# Characteristics of Equilibrated Nonlinear Oscillator Systems

Uri Levy

Weizmann Institute of Science, Rehovot 7610001, Israel
E-mail: uri.levy@weizmann.ac.il

**Abstract**

During the evolution of coupled nonlinear oscillators on a lattice, with dynamics dictated by the discrete nonlinear Schrödinger equation (DNLSE systems), two quantities are conserved: system energy (Hamiltonian) and system density (number of particles). If the number of system oscillators is large enough, a significant portion of the array can be considered to be an "open system", in intimate energy and density contact with a "bath" - the rest of the array. Thus, as indicated in previous works, the grand canonical formulation can be exploited in order to determine equilibrium statistical properties of thermalized DNLSE systems. In this work, given the values of the two conserved quantities, we have calculated the necessary values of the two Lagrange parameters (typically designated $\beta, \mu$) associated with the grand canonical partition function in two different ways. One is numerical and the other is analytic, based on a published approximate entropy expression. In addition we have accessed a purposely-derived approximate PDF expression of site-densities. Applying these mathematical tools we have generated maps of temperatures, chemical potentials, and field correlations for DNLSE systems over the entire thermalization zone of the DNLSE phase diagram, subjected to all system-nonlinearity levels. The end result is a rather complete picture, characterizing equilibrated large DNLSE systems.

## 1. Introduction

The dynamics of a number of periodic systems - molecular, mechanical, optical, and lattice-trapped ultra-cold atoms, can be approximated by the DNLSE. This being the case, a book **[1]**, and a wealth of scientific papers studying the properties of DNLSE-governed systems were published over the years - **[2]**-**[11]** to cite just a few.

The simplest two-term DNLSE reads **[3]**,**[12]**:

$$i \cdot \frac{dU_m}{d\zeta} = C \cdot (U_{m-1} + U_{m+1}) + \gamma \cdot |U_m|^2 \cdot U_m$$

**(1)**

where $\zeta$ is the evolution coordinate (distance or time), $U_m(\zeta)$ is the complex field function of the oscillator at site $m$, the parameter $C$ is the nearest-neighbor coupling constant and $\gamma$ is the unharmonic parameter. In the next section we present a modified dynamics equation (Eq. **(3)**) with the unharmonic parameter $(\gamma)$ replaced by a *nonlinearity parameter* $(\Gamma)$ defined as $\Gamma \equiv \gamma/|C|$.

DNLSE systems are Hamiltonian systems **[1]**,**[13]**. The equivalent Hamiltonian from which the two-term DNLSE is derived (designated $\mathcal{H}_a$ below – Eq. **(5)**) is made up of two "energy" terms **[2]**,**[7]**,**[14]** – a tunneling energy term **[8]** (designated $\mathcal{H}_2$ below – Eq. **(6)**) and interaction energy term **[8]** (designated $\mathcal{H}_4$ below – Eq. **(6)**). The equivalent Hamiltonian (*energy*), is a conserved quantity of the (isolated) system. The other conserved quantity of the system (designated $\mathcal{W}_a$ below – Eq. **(7)**) is *density* (number of particles).

The Hamiltonian-derived DNLSE equation (Eq. **(1)** above or Eq. **(3)** below) has only two integrals of motion (conserved quantities) **[1]**,**[14]**,**[15]** and therefore, for a system of more than two sites, has no general analytic solution. Namely – no analytic expression for amplitude and phase of each and every oscillator at every distance (time) and for every possible set of initial excitation conditions. However, statistical properties of thermalized systems at equilibrium *are* analytically predictable, given the set of initial excitation conditions. Predictions of these equilibrium properties are the subject of the present study.

A DNLSE system initially excited onto a specific point $\left(w_a = \frac{\mathcal{W}_a}{N}, \hbar_a = \frac{\mathcal{H}_a}{N}\right)$ on a well-defined *thermalization zone* of the DNLSE phase diagram (cf. *Figure 2* and Eq. **(2)** below) will thermalize **[7]**. Namely, at long evolution distances all systems initially excited onto the same point of the phase diagram (i.e. same $(w_a, \hbar_a)$ values) are doomed to reach the same equilibrium statistical properties, regardless of initial system-excitation details (cf. *Figure 3* below). And, as stated above, these equilibrium statistical properties are predictable.

Many published papers discuss properties of DNLSE systems initially excited onto the *breather-forming zone* (above the thermalization zone) of the DNLSE phase diagram. Other papers typically discuss a single property of systems initially excited onto the thermalization zone of the DNLSE phase diagram – transition behaviour **[8]**, entropy (with focus on breather dynamics) **[9]**, system instability **[12]**, or system ground states **[16]**. Characteristics of thermalized systems such as PDFs of site-densities (amplitudes squared, designated "$I$" below), temperatures, chemical potentials, and field correlations for all initial excitation conditions and for all values of the nonlinearity parameter ($\Gamma$) were not published in the scientific literature to-date.

Previous studies predicted PDFs and temperatures for strong system nonlinearities based on the quantum phase approximation that justified system-entropy separation into the sum of density-entropy and a relative-phase entropy. Entropy separation led to the derivation of system temperatures as well as analytic expressions of equilibrium PDF($I$) and equilibrium PDF($\theta$) **[17]**-**[19]**.

Here we have taken a rigorous approach of predicting system equilibrium statistics based on the thermodynamics formalism of grand canonical ensembles **[7]** (cf. *Figure 2* below and related text). The thermodynamics approach allowed the extension of system-characteristics predictions to cover the entire thermalization zone – the zone of strong system nonlinearity as well as the zone of weak system nonlinearity.

The grand canonical statistics is determined by two Lagrange parameters - $(\beta, \mu)$. The value of each of these two parameters is uniquely determined by the value of the two DNLSE conserved quantities – density and energy. Once $(\beta(w_a, \hbar_a), \mu(w_a, \hbar_a))$ are determined, equilibrium statistical properties of the system studied such as entropy, temperature, chemical potential, nearest-neighbors field correlations, and PDF($I$) can be calculated.

In the study presented here we have determined $(\beta, \mu)$ given $(w_a, \hbar_a)$ in one of two ways. First, by numerically inverting thermodynamics partial derivatives. Such 2D numerical inversion is challenging and could sometimes introduce uncertainty errors. Second, by finding $\beta_{exp}$ through a direct execution of an analytic expression. We have derived the analytic expression from an approximate analytic expression of system entropy published in **[9]**. Once the value of $\beta_{exp}$ is determined, the value of $\mu_{exp}$ is calculated through solving a *one parameter* implicit integral equation.

In addition to finding equilibrium PDF($I$) by rigorous numerical calculations, we have gone in this study through deriving an approximate analytic PDF($I$) expression. Our derivation here is based on approximating the kernel associated with the partition function in the grand canonical formalism (cf. section **8** and ***Appendix 1***). The easily executed approximate analytic PDF($I$) expression "works" well on a large area of the thermalization zone and brings additional system-characteristics-related insights. For example - the identification of system's temperature with the variance of the system's PDF($I$).

Following is a summarizing list of the mathematical tools we have employed in the present study to *quantitatively* describe key equilibrium properties of thermalized DNLSE systems under all initial excitation conditions:

- Simulations of system evolution: given the amplitude and phase of each and every oscillator in the array, execute the DNLSE (Eq. **(3)** below) to stationary-statistics distances. Next, calculate equilibrium PDFs of densities and of relative phases. Clearly, PDFs calculated this way faithfully describe the PDFs of actual thermalized systems.

- The rigorous grand canonical statistics suggested in **[7]**. This mathematical tool requires numerical inversion of two thermodynamic equations to compute exact values of the two associated Lagrange parameters $(\beta, \mu)$. Small numerical errors in the values determined this way often creep-in.

- The grand canonical statistics with *approximate* values of the two associated Lagrange parameters $(\beta_{exp}, \mu_{exp})$. The parameter $\beta_{exp}$ is determined by an analytic expression derived from an approximate analytic expression for system entropy published in **[9]** (hence the subscript "*exp*", standing for *expression*). The parameter $\mu_{exp}$ is determined next by numerically solving a *single parameter* implicit equation.

- Previously published analytic expressions for PDF(densities) and PDF(relative phase-angles). These analytic expressions hold for high nonlinearity DNLSE systems (strong nonlinearity of the oscillators and/or high average of oscillator amplitudes) **[18]**,**[19]**.

- Approximate analytic expression for PDF(densities). During the present study we have derived an approximate *analytic* expression to PDF(densities) of equilibrated DNLSE systems under a second-order approximation to the kernel of the grand canonical partition function (cf. details in ***Appendix 1***). PDFs calculated through the second-order-derived analytic expression well match the evolution-simulated PDFs for a surprisingly large class of equilibrated DNLSE systems.

In our dynamics-governing DNLSE (Eq. **(3)**), and unlike many DNLSE versions of other studies, we have left the oscillator fields ($U_m's$) dimensional and left the nonlinearity parameter ($\Gamma$) dimensional as well. For example, if the DNLSE system is an array of proximity optical waveguides then the dimension of the complex site functions squared is typically power/volume. In this case the dimension of "energies" is also power/volume. The dimension of the nonlinearity parameter ($\Gamma$), and the dimension of the Lagrange parameter $\beta$ is *inverse*(power/volume). The dimension of Eq. **(3)** below is *sqrt*(power/volume). This amplitude-nonlinearity separation allowed us here the study of the effect of each the three variables - system density ($w_a$), system energy ($\hbar_a$), and the value of the nonlinearity parameter ($\Gamma$) independently, keeping the other two variables constant.

In the next three sections we elaborate on the process of system thermalization and on the related mathematics. In the sections that follow we mathematically analyze and graphically demonstrate the characteristics of equilibrated DNLSE systems: temperature, chemical potential, field correlations and PDF(densities).

Let us start then in discussing the thermalization of DNLSE systems.

## 2. Thermalization of DNLSE systems

In this stage-setting section we describe the DNLSE phase diagram, discuss in more details the grand canonical formulation as applied to DNLSE systems, and present a specific thermalization example. The thermalization example consists of showing evolution-simulated and analytically predicted equilibrium PDFs for two systems. Both systems are placed onto the same location of the phase diagram (same ($w_a, \hbar_a$), see below) but each system is launched by a unique set of initial excitation conditions.

The two conserved quantities, most frequently appearing in this study, are the *intensive site-averaged* quantities. Namely - *system density* ($w_a$) and *system energy* ($\hbar_a$). The two extensive conserved quantities are thus *total system density* - $\mathcal{W}_a = N \cdot w_a$, and *total system energy* - $\mathcal{H}_a = \hbar_a \cdot N$, were $N$ is the number of oscillators in the array.

A DNLSE phase diagram (not to be confused with "phase space") can now be constructed on the ($w_a, \hbar_a$) plane (cf. *Figure 1*).

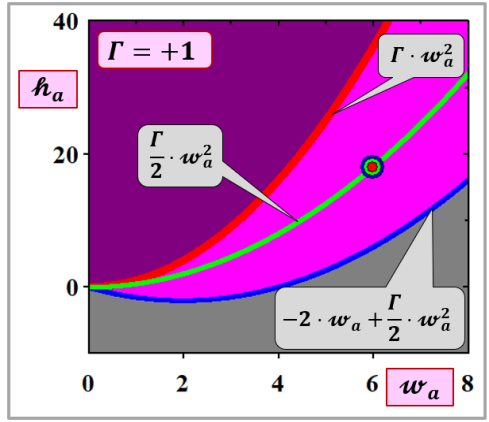

*Figure 1:* *Phase diagram for DNLSE systems with a positive nonlinearity parameter ($\Gamma = +1.0$). The magenta area is the thermalization zone, limited at the bottom by the zero temperature line (or the "ground state line" - $\hbar_{a0}$), and limited at the top by the infinite temperature line ($\hbar_{a\infty}$). Crossing the thermalization zone is an intermediate $L_i$ line. The colored circular disk on the $L_i$ line represents a system with all oscillators initially excited at a uniform amplitude ($= \sqrt{6}$) with random phases in the full range of $[0, 2 \cdot \pi)$. The dark magenta area above the thermalization zone is a negative temperature (or "breather-forming") zone. The study presented here is devoted solely*

*to the equilibrium properties of DNLSE systems initially excited onto the thermalization zone.*

The DNLSE phase diagram is divided into three zones – an inaccessible zone, a thermalization zone, and a negative temperature (or a "breather-forming") zone. The thermalization zone is limited from below by a zero temperature line ($ℏ_{a0}$) and is limited from above by an infinite temperature line ($ℏ_{a∞}$) **[7]**. Crossing the thermalization zone is an intermediate $L_i$ line **[19]**. The mathematical expressions for these three lines are -

$$ℏ_{a0}(w_a) = -2 \cdot w_a + \frac{1}{2} \cdot \Gamma \cdot w_a^2$$
$$L_i(w_a) = \frac{1}{2} \cdot \Gamma \cdot w_a^2$$
$$ℏ_{a∞}(w_a) = \Gamma \cdot w_a^2$$

**(2)**

The $L_i$ line is characterized by two unique properties. The first property is related to system initial excitation conditions. To be on the $L_i$ line, systems can be initially excited with uniform amplitudes and random phases in the full range of $[0, 2 \cdot \pi)$. As the window from which random phase-angles are drawn is getting narrower and narrower, same-amplitude systems are placed closer and closer to the ground state line ($ℏ_{a0}$). To place a system above the $L_i$ line, initial excitation amplitudes must be spread apart, more and more so getting closer and closer to the infinite temperature line ($ℏ_{a∞}$). The second property is related to two limiting cases. The $L_i$ line becomes the lower (upper) limit of the thermalization zone as the nonlinearity parameter ($\Gamma$) goes to infinity (zero).

Following **[19]**, we refer to the thermalization zone below the intermediate $L_i$ line as a *cold zone* and refer to the thermalization zone above the intermediate $L_i$ line as a *hot zone*. In the coming sections we show quantitatively that indeed system temperatures, chemical potentials, field correlations, and PDFs of densities and relative phase-angles, change moderately for systems on the cold zone and change drastically for systems on the hot zone.

The phase diagram of ***Figure 1*** is plotted for a positive nonlinearity parameter ($\Gamma > 0$). In **[16]** the Hamiltonian corresponding to a positive nonlinearity parameter is referred-to as a *positive Hamiltonian*. Here, without loss of generality, we consider only positive Hamiltonian systems, given mirror equivalence. Namely – statistical properties of a system on a point ($w_a, ℏ_a$ ; $\Gamma > 0$) are equal to the properties of a system on its mirror image position - ($w_a, -ℏ_a$ ; $\Gamma < 0$). (In **[8]** the nonlinearity parameter is altogether normalized away, i.e. - its value is fixed at a positive unity).

Schematic of a grand canonical setting as applied to a large array of coupled oscillators is depicted by ***Figure 2***. A large portion of the array – the "open system" is in density contact and in energy contact with the rest of the array – the "bath". At equilibrium, the ensemble averages ($\langle w_a \rangle, \langle ℏ_a \rangle$) of the open system take on the values of the two conserved quantities ($w_a, ℏ_a$) of the bath, determined as soon as the entire array is *initially* excited. Specifically, the values of the Lagrange parameters - $\beta(w_a, ℏ_a), \mu(w_a, ℏ_a)$ of the grand canonical partition function are fixed such that $\langle w_a(\beta, \mu) \rangle = w_a$ and $\langle \hbar_a(\beta, \mu) \rangle = \hbar_a$ [20]. Once $\beta$ and $\mu$ are determined, equilibrium statistical properties of the entire (isolated) array can be analytically predicted.

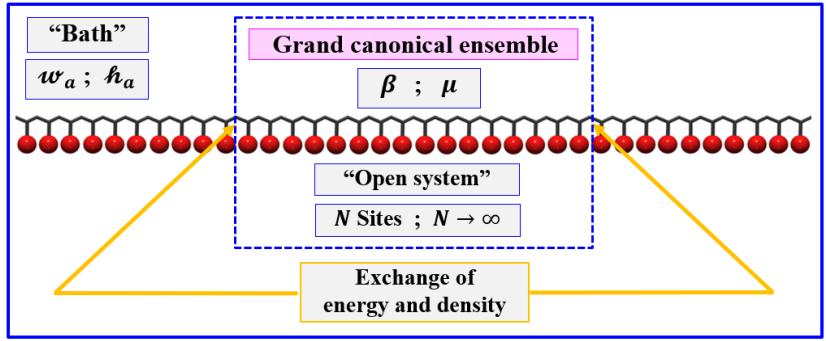

*Figure 2:* *Grand canonical statistics applied to a large array of coupled oscillators. A significant portion of the array is (artificially) assumed to be an "open system" in contact with the rest of the oscillators in the array. At equilibrium, through diffusion-exchange of energy and density between the "bath" and the open system, the grand canonical ensemble averages ($\langle w_a \rangle, \langle \hbar_a \rangle$) settle on values equal to the known conserved density ($w_a$) and conserved energy ($\hbar_a$) values of the entire excited array [20]. This equality dictates the values of the Lagrange parameters - $\beta(w_a, \hbar_a), \mu(w_a, \hbar_a)$. Once $\beta$ and $\mu$ are accurately determined, equilibrium statistical properties of the entire (isolated) array can be faithfully derived.*

Systems on the thermalization zone evolve to equilibrium [7]. Entropy rises to its maximum and stays there [16], and other system statistical properties such as tunneling energy, interaction energy, PDFs of densities and of relative phase-angles, and field correlations, reach a stationary value or shape. These stationary statistical properties are predictable, based on the Gibbsian formalism applied to a grand canonical ensemble [7].

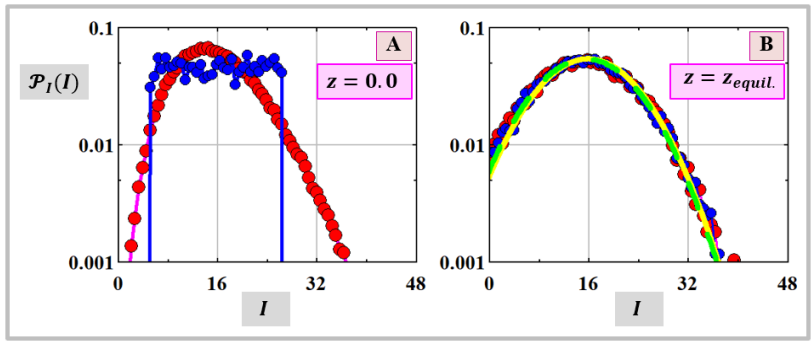

*Figure 3:* *Thermalization of DNLSE systems. A. PDF(I) curves of initial excitation densities of two DNLSE systems. Considering the DNLSE phase diagram, both systems are initially excited onto the **same** thermalization-zone point: $(w_a, \hbar_a) = (16.0, 147.2)$. B. Equilibrium PDF(I) curves of the two same-point-initially-excited systems. The red and blue points are respective PDFs(I) calculated by evolution-to-equilibrium simulations (averaged over six realizations). The shown two practically identical red and blue simulated curves prove system thermalization, independent of*

*initial excitation details. The continuous yellow curve and the dashed green curve are theoretically-predicted equilibrium PDF(I) curves for the given $(w_a, \hbar_a)$ point of the thermalization zone. The yellow curve by previously published analytic expressions assuming strong system nonlinearity* **[19]**. *The dashed green curve by the rigorous grand-canonical statistics as discussed in section* **4** *below.*

Two PDF($I$) predictions of thermalized DNLSE systems are shown in panel B of ***Figure 3***. One curve by previously published analytic expressions assuming strong system nonlinearity **[19]**. Another curve by the rigorous grand-canonical statistics as discussed in section **4** below.

### 3. Dynamics and conserved quantities

The evolution dynamics of a $1d$ array of $N$ coupled unharmonic oscillators is given by Eq. **(1)** above. Throughout this work, without loss of generality, we will use the notation of optics (evolution coordinate $\zeta$ as distance, or $z$ as a normalized distance, with coupled optical waveguides in mind). In this study we assume $N$ to be very large and, subject the equation to periodic boundary conditions ($U_{m+N} = U_m$). The equation consists of two terms – a linear *hopping term*, and a cubic on-site *nonlinear term*.

Several options for normalizing Eq. **(1)** are available **[1]**,**[3]** and are often applied **[10]**,**[17]**. Here, as in **[16]**, we shall eliminate the coupling constant from Eq. **(1)** except for its sign, following a division by $|C|$ –

$$i \cdot \frac{dU_m}{dz} = signC \cdot (U_{m-1} + U_{m+1}) + \Gamma \cdot |U_m|^2 \cdot U_m$$

$$z \equiv |C| \cdot \zeta \; ; \; signC \equiv sign(C) \; ; \; \Gamma \equiv \frac{\gamma}{|C|}$$

**(3)**

In Eq. **(3)**, the amplitudes $U'_m s$ are dimensional. The evolution coordinate ($z$) – "distance" (or "time") is dimensionless. The dimension of the nonlinearity parameter $\Gamma$ is $[U_m]^{-2}$. *System nonlinearity* is expressed by the dimensionless product of the nonlinearity parameter and the conserved system density ($i.e. as (\Gamma \cdot w_a)$). If $signC = sign(\Gamma) / -sign(\Gamma)$ then Eq. **(3)** is a "focusing" / "defocusing" version of the DNLSE systems **[1]**,**[21]**.

It is convenient at this point, and indeed done in almost every DNLSE study, to perform a Madelung transformation on the complex canonical coordinates $(U_m, i \cdot U_m^*)$ into the set of density-angle real canonical polar coordinates: $(U_m, i \cdot U_m^*) \rightarrow (I_m, \phi_m)$. The complex field functions $(U_m's(z))$ of Eq. **(3)** take then the form:

$$U_m = u_m \cdot e^{i \cdot \phi_m} \; ; \; u_m \equiv \sqrt{I_m} \; ; \; \theta_m \equiv \phi_m - \phi_{m+1}$$

**(4)**

Equation **(3)** can be derived from an equivalent Hamiltonian ($\mathcal{H}_a$) which is a conserved quantity, associated with the system's time translation invariance **[13]**,**[22]** -

$$\mathcal{H}_a = \sum_{m=1}^{N} \left\{ signC \cdot (U_m^* \cdot U_{m+1} + U_m \cdot U_{m+1}^*) + \frac{\Gamma}{2} \cdot |U_m|^4 \right\}$$

**(5)**

The variables $(U_m, i \cdot U_m^*)$ are canonical variables. Adopting the assignment $q_m = U_m$ ; $p_m = i \cdot U_m^*$, Eq. **(3)** is derived from the Hamiltonian **(5)** as $\frac{dU_m}{dz} = \frac{\partial \mathcal{H}_a}{\partial (i \cdot U_m^*)}$.

In the polar variables $(I_m, \phi_m)$ of Eq. **(4)**, the DNLSE Hamiltonian (Eq. **(5)**) takes on the form:

$$\mathcal{H}_2(z) = signC \cdot \sum_{m=1}^{N} 2 \cdot u_m u_{m+1} \cdot \cos \theta_m \quad ; \quad \mathcal{H}_4(z) = \sum_{m=1}^{N} \frac{\Gamma}{2} \cdot u_m^4$$

$$\mathcal{H}_a \equiv \mathcal{H}_2(z) + \mathcal{H}_4(z)$$

$$\hbar_2(z) \equiv \frac{\mathcal{H}_2(z)}{N} \quad ; \quad \hbar_4(z) \equiv \frac{\mathcal{H}_4(z)}{N} \quad ; \quad \hbar_a \equiv \frac{\mathcal{H}_a}{N}$$

**(6)**

The Hamiltonian energies - $(\mathcal{H}_2(z), \mathcal{H}_4(z))$ are the nearest-neighbor *tunneling energy* term and the on-site *interaction energy* term respectively. Obviously, both $\mathcal{H}_2(z)$ and $\mathcal{H}_4(z)$ vary with propagation distance, but their sum does not. During DNLSE evolution then, an energy diffusion process transfers energy from $\mathcal{H}_2(z)$ to $\mathcal{H}_4(z)$ or the other way around.

Another conserved quantity of DNLSE systems, thanks to the system's invariance with respect to global phase rotations **[1]**,**[13]**, is "density" $(\mathcal{W}_a)$ (or *norm*, or *number of particles*) given by -

$$\mathcal{W}_a = \sum_{m=1}^{N} I_m(z) \quad ; \quad w_a \equiv \frac{\mathcal{W}_a}{N}$$

**(7)**

As stated above, system density and system energy $(w_a, \hbar_a)$ form a plane over which the DNLSE phase diagram is graphically represented (cf. ***Figure 1***).

## 4. Grand canonical partition function and PDF of site-densities

The grand canonical partition function $(\mathcal{Z}(\beta, \mu))$ is given as **[7]**:

$$\mathcal{Z}(\beta, \mu) = \int_0^{\infty} \int_0^{2\pi} \prod_{m=1}^{N} d\phi_m \cdot dI_m \cdot e^{-\beta \cdot (\mathcal{H} + \mu \cdot \mathcal{W})}$$

**(8)**

Following integration over the phase variable ($\phi_m$) the expression for the partition function is reduced to -

$$Z(\beta,\mu) = (2\pi)^N \cdot \int_0^\infty \prod_{m=1}^N \mathcal{K}(I_m, I_{m+1}) \cdot dI_m$$

$$\mathcal{K}(I_m, I_{m+1}) \equiv e^{\left[-\beta \cdot \left(\frac{\mu}{2} \cdot I_m + \frac{\Gamma}{4} \cdot I_m^2\right) - \beta \cdot \left(\frac{\mu}{2} \cdot I_{m+1} + \frac{\Gamma}{4} \cdot I_{m+1}^2\right)\right]} \cdot \mathcal{I}_0\left(2 \cdot \beta \cdot \sqrt{I_m \cdot I_{m+1}}\right)$$

(9)

where $\mathcal{I}_0(\cdot)$ is the zero order of modified Bessel function of the first kind, satisfying for integer $n$ [23]: $\mathcal{I}_n(z) = \frac{1}{\pi} \cdot \int_0^\pi e^{z \cdot \cos\theta} \cdot \cos(n \cdot \theta) \cdot d\theta$. The nonnegative symmetric kernel $(\mathcal{K}(x,y))$ of Eq. (9) satisfies $\int \mathcal{K}(x,y) \cdot v_i(x) \cdot dx = \lambda_i \cdot v_i(y)$. If the array of oscillators is large enough then the partition function, to a very good approximation, is given by the largest eigenvalue of the kernel [7]:

$$Z(\beta,\mu) \cong [2 \cdot \pi \cdot \lambda_1(\beta,\mu)]^N$$

(10)

and the PDF($I$) of site-densities $(\mathcal{P}_I(I))$ is given by the square of the eigenfunction $(v_1(I))$ associated with the largest eigenvalue [7]:

$$\mathcal{P}_I(I) = v_1^2(I)$$

(11)

For numerical calculations we treat the kernel ($\mathcal{K}(x,y)$) as a 2D matrix on a $\Delta$-spaced grid and solve for $\lambda_i$ and $v_i$:

$$(\mathcal{K}(x,y) \cdot \Delta) \cdot v_i = \lambda_i \cdot v_i$$

(12)

where now $v_i$ are the normalized eigenvectors of the matrix $(\mathcal{K}(x,y) \cdot \Delta)$.

The kernel $\mathcal{K}(x,y)$ is written explicitly as:

$$\mathcal{K}(x,y) = e^{\left[-\beta \cdot \left(\frac{\mu}{2} \cdot x + \frac{\Gamma}{4} \cdot x^2\right) - \beta \cdot \left(\frac{\mu}{2} \cdot y + \frac{\Gamma}{4} \cdot y^2\right)\right]} \cdot \mathcal{I}_0\left(2 \cdot \beta \cdot \sqrt{x \cdot y}\right)$$

(13)

Two sets of equations relate the two conserved quantities $(w_a, \hbar_a)$ to the two Lagrange parameters of the grand canonical partition function $(\beta, \mu)$. The first set is the average-calculating integrals –

$$w_a = \langle w_a(\beta,\mu) \rangle = \int_0^\infty I \cdot v_1^2(\beta,\mu; I) \cdot dI$$

$$\hbar_a = \langle \hbar_a(\beta,\mu) \rangle = \frac{\langle \mathcal{H}_a(\beta,\mu) \rangle}{N}$$

(14)

The second set is the set of partial derivatives of the grand canonical partition function [20]:

$$w_a = \langle w_a(\beta,\mu) \rangle = -\frac{1}{\beta} \cdot \frac{\partial\{ln[\lambda_1(\beta,\mu)]\}}{\partial \mu}$$

$$\hbar_a = \langle \hbar_a(\beta,\mu) \rangle = -\frac{\partial\{ln[\lambda_1(\beta,\mu)]\}}{\partial \beta} - \mu \cdot w_a(\beta,\mu)$$

(15)

In equations (14) and (15), $v_1^2(\beta,\mu;I)$, $\langle \mathcal{H}_a(\beta,\mu) \rangle$, and $\lambda_1(\beta,\mu)$ are each a 2D map calculated through Eqs. (12) and (11). Inverting (numerically) one of these two important equation sets yields the exact value of $\beta(w_a, \hbar_a), \mu(w_a, \hbar_a)$. Once $(\beta, \mu)$ are uniquely determined, exact equilibrium statistical properties of a system excited onto a point $(w_a, \hbar_a)$ of the thermalization zone can be determined.

In the following sections we present maps of temperatures, chemical potentials, and field correlations, for equilibrated DNLSE systems. Map calculations cover the entire thermalization zone and hold for all system nonlinearity levels. The maps reveal, for example, the universality property of field correlations, and illuminate the inverse relation between system temperatures and field correlations. Further, we present here-derived analytic expressions (cf. *Appendix 1*) that well approximate the grand canonical partition function and the PDF of site-densities on a large portion of the thermalization zone.

We start with the temperature of a DNLSE system.

## 5. Temperature of a DNLSE system

This temperature section consists of three parts. In the first part we rigorously derive an expression for the temperature of a DNLSE system. In the second part we present an analytic expression for the temperature following an entropy expression published in [9]. The third part is devoted to graphical illustrations.

*Exact expression*. Temperature of an equilibrated DNLSE system is calculated through the energy derivative of Gibbs entropy. The Gibbs entropy equation relates the entropy of a system to the probability distribution of the microstates [24]. Applied to the grand canonical ensemble considered here, continuous Gibbs entropy is given as [8],[17]:

$$S(\beta,\mu) \equiv -k \cdot \int_0^\infty \int_0^{2\pi} \mathcal{P}(\beta,\mu) \cdot ln[h \cdot \mathcal{P}(\beta,\mu)] \, d^N I \cdot d^N \theta$$

(16)

where $k$ is the Boltzmann constant (later on to be set at dimensionless unity). The integration sign $d^N I$ stands for $\prod_1^N dI_m$ and similarly for $d^N \theta$. The value of the scale parameter $h$, with units of $energy^N$, is set at one.

The probability density of any given state $\mathcal{P}(I_1, I_2, \ldots I_N, \theta_1, \theta_2, \ldots \theta_N; \beta, \mu)$ is

$$\mathcal{P}(I_1, I_2, \ldots I_N, \theta_1, \theta_2, \ldots \theta_N; \beta, \mu) = \frac{1}{\mathcal{Z}(\beta, \mu)} \cdot e^{-\beta \cdot (\mathcal{H} + \mu \cdot \mathcal{W})}$$

(17)

Note the use of the pair $(I_m, \theta_m)$ (cf. Eq. **(4)**) in the entropy expression **(16)** with the probability density **(17)** **[8]**,**[17]** and not the use of the canonical pair $(I_m, \phi_m)$.

Inserting **(17)** into **(16)** and working out the integral using **(10)**:

$$\frac{1}{k} \cdot S(\beta, \mu) = N \cdot \ln[2 \cdot \pi \cdot \lambda_1(\beta, \mu)] + \beta \cdot N \cdot \langle \hbar_a + \mu \cdot w_a \rangle$$

(18)

Defining site-averaged system entropy $s = S/N$, setting $k = 1$, and using $\langle w_a(\beta, \mu)\rangle = w_a$ ; $\langle \hbar_a(\beta, \mu)\rangle = \hbar_a$, we arrive at:

$$s(\beta, \mu) = \ln[2 \cdot \pi \cdot \lambda_1(\beta, \mu)] + \beta \cdot (\hbar_a + \mu \cdot w_a)$$

(19)

Since the Lagrange multipliers $(\beta, \mu)$ are each a function of $(w_a, \hbar_a)$, Eq. **(19)** for the site-averaged system entropy $(s)$ rigorously reads:

$$s(w_a, \hbar_a) = \ln[2 \cdot \pi \cdot \lambda_1(w_a, \hbar_a)] + \beta(w_a, \hbar_a) \cdot [\hbar_a + \mu(w_a, \hbar_a) \cdot w_a]$$

(20)

In the present study we adhere to the definition of DNLSE system temperature as put forward in **[19]**:

$$T_{DNLSE}(w_a, \hbar_a) \equiv \left( \Gamma \frac{\partial s(w_a, \hbar_a)}{\partial \hbar_a} \right)^{-1}_{w_a}$$

(21)

Taking the derivative of $s(w_a, \hbar_a)$ given by Eq. **(20)** (multiply by $\Gamma$ and invert later), we arrive at the following three term expression (keeping $w_a$ fixed):

$$\frac{\partial s(\hbar_a)}{\partial \hbar_a} = \frac{\partial \ln[\lambda_1(\hbar_a)]}{\partial \hbar_a} + w_a \cdot \frac{\partial [\beta(\hbar_a) \cdot \mu(\hbar_a)]}{\partial \hbar_a} + \beta(\hbar_a)$$

(22)

Somewhat surprisingly, as we have shown numerically, for all $(w_a, \hbar_a)$ points of the thermalization zone of the DNLSE phase diagram, the first two terms of Eq. **(22)** cancel out:

$$\frac{\partial \ln[\lambda_1(\hbar_a)]}{\partial \hbar_a} + w_a \cdot \frac{\partial [\beta(\hbar_a) \cdot \mu(\hbar_a)]}{\partial \hbar_a} = 0$$

(23)

So that after multiplying by $\Gamma$ and inverting we are left with

$$T_{DNLSE}(w_a, \hbar_a) \equiv \left( \Gamma \frac{\partial s(w_a, \hbar_a)}{\partial \hbar_a} \right)^{-1}_{w_a} = \frac{1}{\beta(w_a, \hbar_a) \cdot \Gamma}$$

(24)

Down below we use just "$T$" for the exact $T_{DNLSE}$ given by Eq. (24).

The dimension of the Lagrange parameter $\beta$ is the inverse of the equivalent-Hamiltonian dimension, i.e. $energy^{-1}$. Similarly, the dimension of the nonlinearity parameter $\Gamma$ is also $energy^{-1}$. The dimension of the temperature of a DNLSE system is thus $energy^2$. Note that $energy^2$ is also the dimension of the variance of PDF($I$).

*Analytic expression*. The author of **[9]** derived an expression for system entropy that is an approximation in the center of the thermalization region, and is exact at the margins. Modifying the derived expression (to explicitly include the nonlinearity parameter ($\Gamma$)) and taking the energy derivative (at constant density - Eq. (21)) we arrive at the analytic expression –

$$T_{exp}(w_a, \hbar_a) = \frac{1}{\Gamma} \cdot \frac{4 \cdot \left( w_a + \frac{\hbar_{a,max}(w_a)}{4} \right)^2 - \Delta \hbar_a^2(w_a, \hbar_a)}{2 \cdot \Delta \hbar_a(w_a, \hbar_a)}$$

$$\hbar_{a,max}(w_a) \equiv \Gamma \cdot w_a^2 \ ; \ \Delta \hbar_a(w_a, \hbar_a) \equiv \hbar_{a,max}(w_a) - \hbar_a$$

(25)

From Eqs. (24) and (25) the Lagrange parameter $\beta_{exp}$ is determined and then the other Lagrange parameter - $\mu_{exp}$, is calculated through solving a *one parameter* implicit equation given by the integral (14):

$$\langle w_a(\beta_{exp}, \mu_{exp}) \rangle - w_a = \int_0^\infty I \cdot v_1^2(\beta_{exp}, \mu_{exp}; I) \cdot dI - w_a = 0$$

(26)

Once $(\beta_{exp}, \mu_{exp})$ are determined, system PDF($I$) can be numerically computed through Eq. (12).

On the $L_i$ line (Eq. (2)), Eq. (25) is reduced to a particularly simple expression -

$$T_{exp,L_i}(w_a) = \frac{4 + 2 \cdot w_a \cdot \Gamma}{\Gamma^2}$$

(27)

Thus, if $2 \cdot w_a \cdot \Gamma \gg 4$ then the temperature along the $L_i$ line is the straight line $2 \cdot w_a/\Gamma$, in agreement with **[19]**. If, on the other hand, $2 \cdot w_a \cdot \Gamma \ll 4$ then the temperature along the $L_i$ line is independent of system density ($w_a$) and goes to infinity with $\Gamma \to 0$ as $\Gamma^{-2}$.

Consulting Eq. (25), it is clear that in general (not only on the $L_i$ line) as $\Gamma \to 0$ temperatures of systems on all points of the thermalization zone (except for systems in the ground state) go to infinity. PDFs($I$) on the other hand are still of a finite width (see for example panel A of *Figure 17* below).

Two mathematical methods are at our disposal then to calculate the temperature of a DNLSE system initially excited onto a $(w_a, \hbar_a)$ point on the DNLSE phase diagram. One method – Eq. **(24)** -involves numerical calculation of a $\lambda_1(\beta, \mu)$ map (Eq. **(12)**) and finding $(\beta, \mu)$ to satisfy the partial derivatives of Eq. **(15)**. Mathematically this numerical procedure should yield the exact temperature (Eq. **(24)**). Practically, a certain uncertainty in the determined value creeps-in as the vicinity of the minimum of the maps involved is rather shallow. The other method is a direct execution of the analytic expression **(25)**. In the graphical illustrations below we use both ways as the case may be.

*Graphical illustrations*. Temperature characteristics of equilibrated DNLSE systems such as temperature dependence on system energy (exponential-like), or dependence of temperature on the value of the nonlinearity coefficient (inverse relations) are illustrated by *Figure 4* through *Figure 10*.

In *Figure 4* we show that temperatures calculated analytically (Eq. **(25)**) reasonably approximate the more rigorously determined system temperatures (inversion of expressions **(15)** to determine $\beta(w_a, \hbar_a)$).

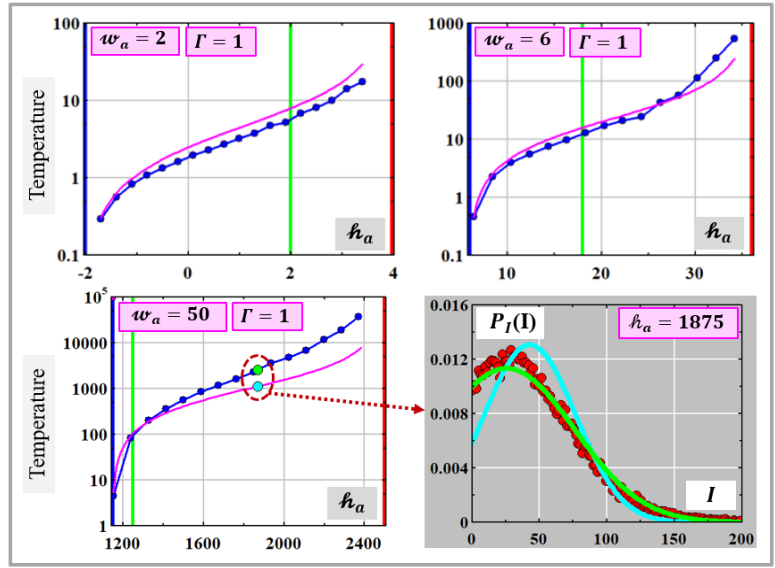

*Figure 4: Temperatures of equilibrated DNLDE systems. Three of the four panels show temperatures on a log scale vs. system energy at a fixed system density as marked. Two curves are shown on each of the three panels. The discrete blue dots on the blue curve, showing exact temperatures, were calculated through inversion of the partial derivatives of Eq. (15). The magenta curves were calculated by the analytic expression (25). The blue-green-red vertical lines designate the crossings of the $\hbar_{a0}$; $L_i$; $\hbar_{a\infty}$ lines respectively (cf. Eq. (2)). The lower-right panel shows three curves of $\mathcal{P}_I(I)$ corresponding to the two marked points on the lower left panel. The red-points curve is the result of simulations, evolving the system to equilibrium (averaged over sixteen realizations). The green curve, pretty much following the red-points curve, corresponds to the green point on the left panel and was calculated through the exact procedure (Eq. (15) and Eq. (12)). The cyan curve corresponds to the cyan dot on the left panel and was calculated by the analytic expression (Eq. (25)). Generally, the figure indicates that whereas the temperature values predicted analytically are approximate, their general dependence on system energy and system density follows the exact-temperature dependence.*

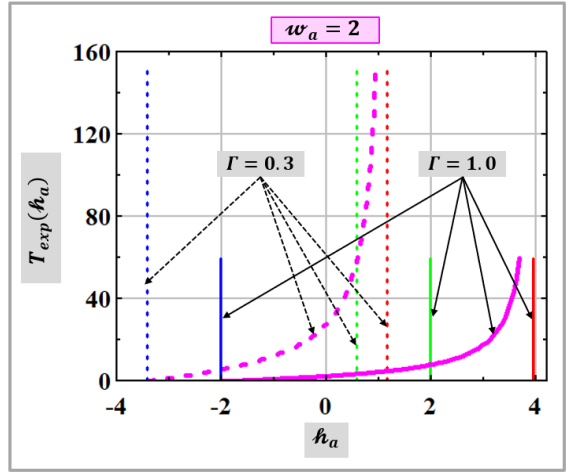

*Figure 5: Temperature vs. system energy (Eq. (25)) at a fixed system density ($w_a = 2.0$), for two values of the nonlinearity parameter ($\Gamma = 0.3 \,; 1.0$). Energy range for both $\Gamma$ values is $\hbar_{a0} + 0.01 \cdot \Delta\hbar_a$ to $\hbar_{a0} + 0.95 \cdot \Delta\hbar_a$. Vertical lines as in Figure 4. Note the very high system temperatures at the low $\Gamma$ value.*

The curves of *Figure 5* show the rise of temperatures with system energy at a fixed system density (travelling vertically on the phase diagram). Initially temperatures rise linearly with energy, going to kind of an exponential rise to infinity for systems initially excited with higher and higher energies. Note the temperature "restriction" effect of increased system nonlinearity ($\Gamma \cdot w_a$).

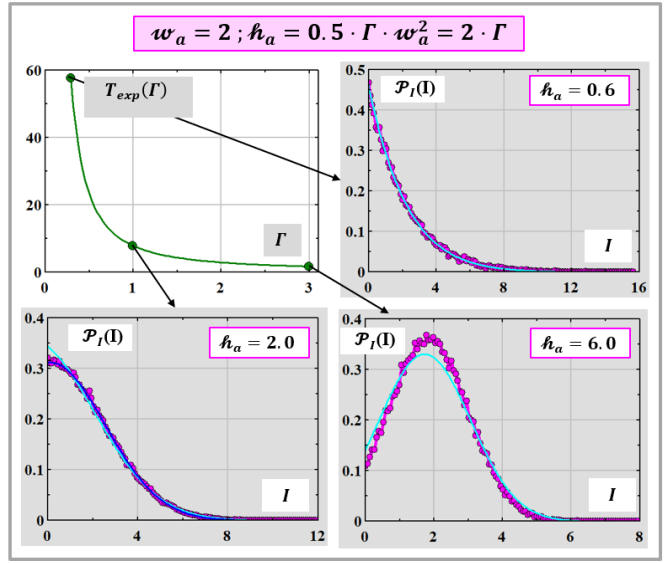

*Figure 6: Temperatures and PDFs(I) for a system on the $L_i$ line having a fixed, relatively low, system density ($w_a = 2.0$). Temperatures and PDFs(I) are shown for three different values of the nonlinearity parameter ($\Gamma = 3.0 \,; 1.0 \,; 0.3$). Top left panel shows analytically calculated system temperature: $T_{exp} = 1.78 \,; 8.0 \,; 57.8$ respectively. The other three panels show simulated $P_I(I)$ curves by magenta dots corresponding to the three values of the nonlinearity parameter as indicated by the black arrows. The cyan curves were numerically calculated (Eq. (12)) given*

$(\beta_{exp}, \mu_{exp})$ *(Eqs.* **(24)**-**(26)***). The dark blue curve on the lower left panel is an exact PDF(I) (Eqs.* **(15)**, *and* **(12)***). The figure shows the very strong dependence of system temperature on the value of the nonlinearity parameter $\Gamma$, for a system on the $L_i$ line, characterized by a relatively low system density of $w_a = 2.0$.*

The next two figures - ***Figure 6*** and ***Figure 7*** show equilibrium temperatures for systems initially excited (at $z = 0$) onto the intermediate ($L_i$) line: uniform amplitudes, $u_m = \sqrt{2}$ for all $m$, and random phases in the full range of $[0,2 \cdot \pi)$. The figures show the sharp rise of system temperature as the value of the nonlinearity parameter ($\Gamma$) is reduced. Note that as $\Gamma \to 0$ the infinite temperature line ($h_{a\infty}$) gets closer and closer to the intermediate $L_i$ line (where the system in question resides), until the two lines merge at $\Gamma = 0$ and system temperature is infinitely high (***Figure 7***). It is fair then to reach the very intuitive conclusion that the energy-distance of a system from the $h_{a\infty}$ line is a strong indication to its temperature: shorter distance – higher temperature.

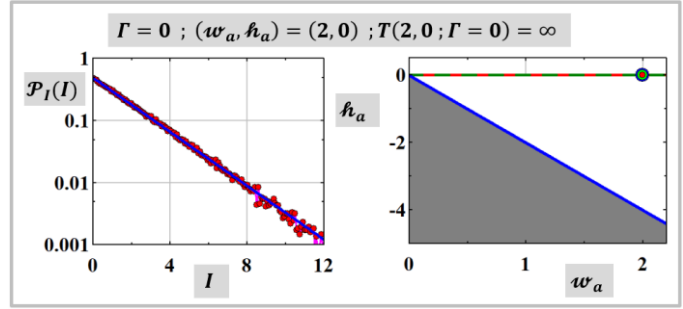

***Figure 7:*** *Equilibrium distribution of densities $\left(\mathcal{P}_I(I)\right)$ for a system on the $L_i$ line (with $w_a = 2.0$) at zero system nonlinearity ($\Gamma \cdot w_a = 0$). The left-panel-shown densities distribution $\left(\mathcal{P}_I(I)\right)$ of the oscillator array is a decaying exponent* **[17]**, *corresponding to an infinite system temperature (cf. Eqs.* **(25)** *and* **(27)***). The right panel shows the position of the system on the DNLSE phase diagram. Note the merging of the $\hbar_{a\infty}(w_a)$ line with the $L_i(w_a)$ line (Eq.* **(2)***).*

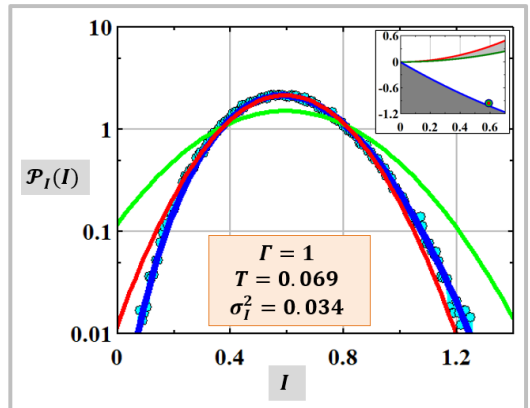

***Figure 8:*** *Equilibrium distribution of site-densities for a weak system nonlinearity ($\Gamma \cdot w_a = 0.6$) and a very cold system (very close to the ground state line, see the inset). The simulated $\mathcal{P}_I(I)$ is shown by the cyan-colored dots. Three continuous colored curves are overlaid. Blue – the exact theoretically predicted $\mathcal{P}_I(I)$ curve for the specific point on the phase diagram $[(w_a, \hbar_a) = (0.6, -0.95)]$ at $\Gamma = 1$. Red – a*

*perfect-Gaussian least-mean-squared fitted to the blue curve. The figure shows then that the equilibrium PDF(I) is NOT everywhere on the phase diagram a perfect, possibly-truncated, Gaussian. Or, to make a positive statement, at weak nonlinearities and low temperatures, the shape of the distribution of DNLSE site-densities may deviate from a perfect possibly-truncated Gaussian. At strong nonlinearities, the equilibrium distribution of site-densities everywhere on the thermalization zone of the DNLSE phase diagram is a pure possibly-truncated Gaussian ([19] and section 8). Light green – a Gaussian with variance equals to the temperature of the system $\{T_{DNLSE}(w_a, \hbar_a) = (\beta(w_a, \hbar_a) \cdot \Gamma)^{-1}$, Eq. (24)$\}$. At weak nonlinearities then, system temperature may be significantly larger than the `variance` of the distribution of site-densities (see also Figure 9).*

At weak system nonlinearities, the shape of PDFs(I) deviate from a pure Gaussian shape. In addition, the value of system temperature no longer coincides with the `variance` (the variance of an untruncated Gaussian) of the PDF(I), but rather deviates to the high side. This is shown by *Figure 8* and by *Figure 9*.

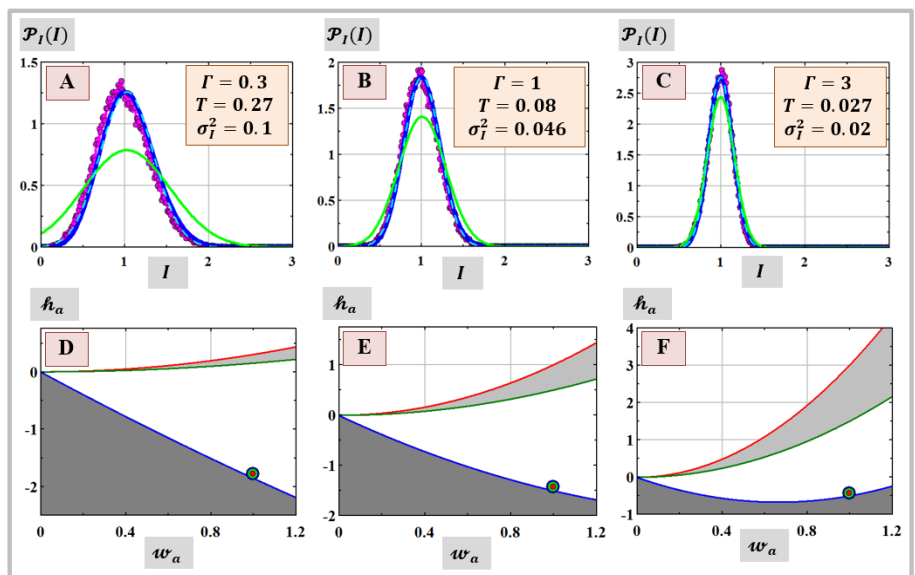

*Figure 9: Equilibrium distributions of site-densities ($\mathcal{P}_I(I)$) for weak nonlinearities and very cold systems, all at system density of $w_a = 1$. Value of the nonlinearity coefficient for [(A,D),(B,E),(C,F)] is sequentially increased as $\Gamma = (0.3, 1.0, 3.0)$ respectively. Each panel of the top row shows four curves: magenta – simulated-to-equilibrium PDF(I). Blue – exact theoretically predicted equilibrium PDF(I). Dashed light-blue – fit of a Gaussian curve to the theoretical PDF(I) curve (very small deviation in A and close match in B and C). Light green - a Gaussian with variance equal to the temperature of the system (Eq. (24)). Comparing the green curves to the other curves of each panel, going from A to C, we see how the large difference of temperature vs. variance of the PDF(I) at low system nonlinearity gradually shrinks as the value of the nonlinearity coefficient is increased. The panels of the bottom row show the position of the analyzed systems on the respective phase diagrams.*

The last figure of this temperature section - *Figure 10* – presents two maps of log-temperatures for DNLSE systems on the low-nonlinearity portion of the thermalization zone ($\Gamma \cdot w_a$ of order unity). Shown temperatures were calculated analytically (Eq. **(25)**). The two maps presented here are complementary to a temperature map for a stronger nonlinearity portion of the thermalization zone, presented in an already-published study **[19]**.

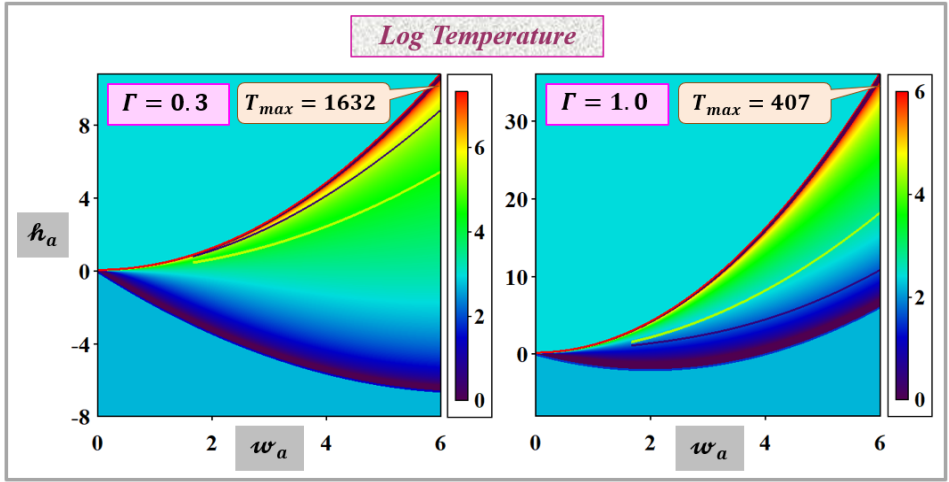

*Figure 10: Maps of log temperatures on a weak nonlinearity portion (system nonlinearity of order unity) of the thermalization zone. Left: $\Gamma = 0.3$. Right: $\Gamma = 1.0$. The maps were calculated analytically (Eq. **(25)**). The yellow lines crossing a green area on both maps mark the respective $L_i$ lines. The dark lines crossing along the top of the green area on the left map and crossing the blue area on the right map are isotherm lines, having – as shown – a concave shape. These low nonlinearity maps complement a temperature map for a stronger nonlinearity portion (system nonlinearity greater then order unity) of the thermalization zone presented in an already-published study **[19]**.*

With *Figure 10* our discussion of system temperatures is concluded. We proceed now to discuss the associated chemical potentials of these equilibrium-reached DNLSE systems.

## 6. Chemical potentials of equilibrium-reached DNLSE systems

The multiplier $\mu$ in the grand canonical partition function (Eq. **(8)**) is introduced in analogy with a chemical potential to ensure conservation of total system density ($\mathcal{W}_a$) **[7]**. Note that here, different from the standard thermodynamic chemical potential that has a dimension of physical energy **[25]**, the equivalent chemical potential introduced in Eq. **(8)** is dimensionless.

Back to conservation of total system density - indeed, we calculate the value of the density-conserving chemical potential $\left(\mu_{exp}(w_a, \hbar_a)\right)$ for a system at $(w_a, \hbar_a)$ by solving the *one parameter* implicit integral equation **(26)** to equate the known system density ($w_a$) with its computed average ($\langle w_a \rangle$). Equation **(26)** is solved with a known $\beta_{exp}$, as determined by the analytic expression **(25)** (and applying Eq. **(24)**). We thus designate the so-calculated chemical potential by the subscript "$exp$".

The next three figures highlight key characteristics of the chemical potential of systems initially excited onto the thermalization zone of the DNLSE phase diagram and evolved to equilibrium.

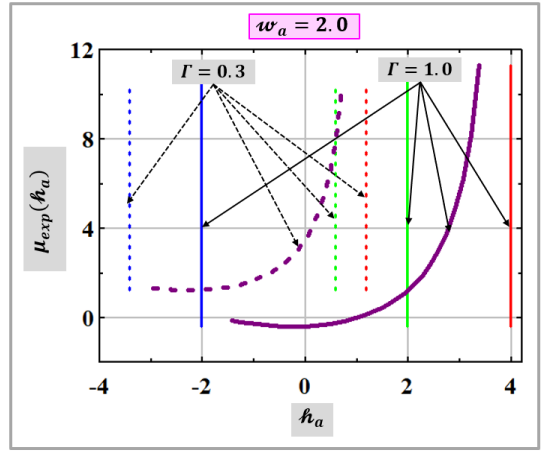

*Figure 11: Chemical potential $(\mu_{exp})$ as a function of system energy for a fixed system density ($w_a = 2.0$). Colored vertical lines as in **Figure 4**. Two curves are shown, corresponding to two values of the nonlinearity parameter as marked. Somewhat like temperature, values of the chemical potential rise sharply as system energy $(\hbar_a)$ gets closer to the infinite temperature line $(\hbar_{a\infty})$. Like the value of system temperature, the value of the chemical potential vary strongly with the value of the nonlinearity parameter $(\Gamma)$. Interestingly, unlike temperature curves, the rise of the chemical potential curve with system energy is not monotonic. At small system energies, close to the ground state energy, chemical potential values are shown to decrease with system energy before the onset of a monotonic rise.*

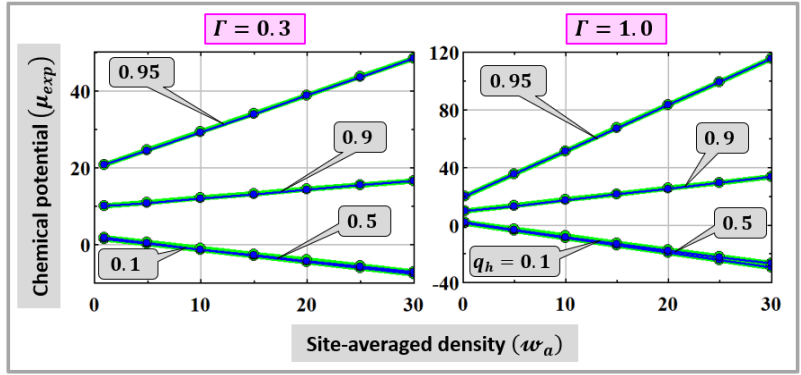

*Figure 12: Chemical potential $(\mu_{exp})$ as a function of site-averaged system density for four $q_h$-values (the parameter $q_h$ designates a fraction of the energy span - $\Delta\hbar_a$, i.e.: $\hbar_a(w_a) = \hbar_{a0}(w_a) + q_h \cdot \Delta\hbar_a(w_a)$). Bottom to top - ($q_h = 0.1, 0.5, 0.9, 0.95$) as marked. Left: $\Gamma = 0.3$. Right: $\Gamma = 1.0$. The blue dots were numerically calculated by solving the one parameter implicit integral equation **(26)**). The fitted straight green lines, each coincides with the first and last data point of the respective $q_h$ curve. It is quite illuminating to realize that for a fixed $q_h$ value, the values of the chemical potentials of DNLSE systems at equilibrium essentially fall on a straight line.*

The curves of *Figure 11* show the dependence of equilibrium chemical potential on system energy for fixed system density as marked (going up vertically on the phase diagram). The two sets of curves in each of the panels were calculated for two different values of the nonlinearity parameter ($\Gamma = 0.3\,;1.0$). Moving up in energies, the curves show fast rise of chemical potential values for high system energies, but, interestingly, show slow *fall* of values for system energies close to the ground state energy ($\hbar_{a0}$).

The two panels of *Figure 12* reveal an interesting dependence of the DNLSE chemical potential on system density. The panels show that chemical potential values plotted against system density for systems with energies at a fixed ratio ($q_h$) of the total energy span $\left(\hbar_a(w_a) = \hbar_{a0}(w_a) + q_h \cdot \Delta h_a(w_a)\right)$ form a straight line. Not surprising, starting value (for a near-zero density) and slope of a straight line depend on the value of the nonlinearity parameter (cf. left panel vs. right panel). Such regular dependence could provide a clue for deriving an independent analytic expression to calculate equilibrium chemical potentials of DNLSE systems (not done in the present study).

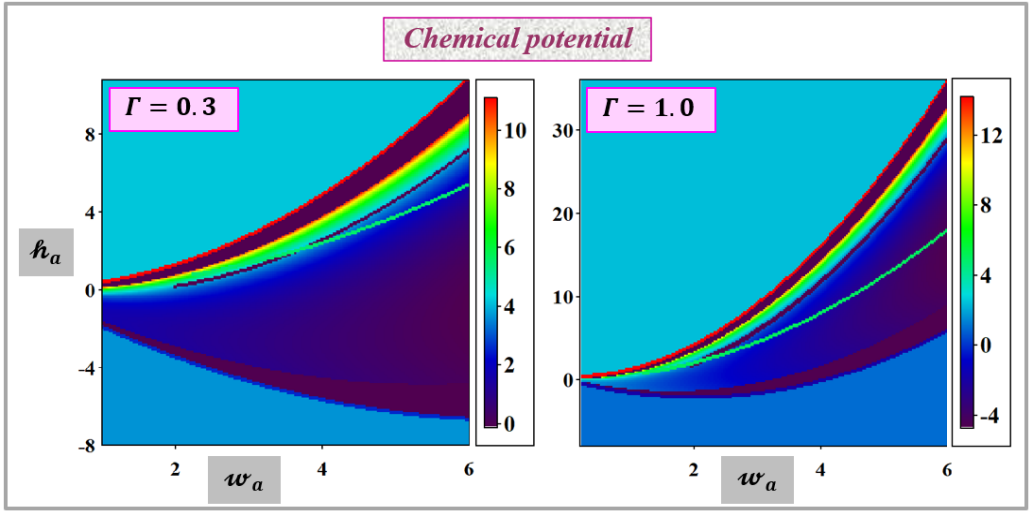

*Figure 13:* *Maps of chemical potential $(\mu_{exp})$ shown on the thermalization zone of the DNLSE phase diagram. The green line crossing the map designates the $L_i$ line. The concave dark line crossing each of the maps is an equipotential line.*

The last figure of this section - *Figure 13*, presents two maps of DNLSE chemical potentials. Note that, different from the temperature maps (*Figure 10*), the color scale is linear. Looking at the maps, we see that values of DNLSE chemical potentials, like temperature values, sharply rise with energies if the systems are initially excited onto a high energy region of the thermalization zone.

With *Figure 13* our discussion of chemical potentials is concluded. We proceed now to discuss the important dynamic phenomenon of evolving DNLSE systems – field correlation.

## 7. Field correlations of DNLSE systems

A key characteristic of evolving DNLSE systems is the formation of field correlations. The formation of correlations is perhaps best viewed in terms of energy diffusion between interaction energy and tunneling energy, keeping the sum constant. Consider for example a system initially excited onto the $L_i$ line with uniform amplitudes, say $u_m = u_0$ for all $m$ and random phases in the full range of $[0, 2 \cdot \pi)$. On system launch, the tunneling energy is zero - $\hbar_2(0) = 0$, and interaction energy is at a minimum of $\hbar_4(0) = \frac{1}{2} \cdot \Gamma \cdot u_0^4$. As the system evolves, amplitudes spread, amplitude-related entropy is generated, and the interaction energy goes to higher values. Necessarily, to keep the sum of energies constant, tunneling energy must go to negative values, forcing the relative phases of neighboring sites to pile-up close to $\pi$ (to get negative $\cos \theta's$ as signC is positive) and thus field correlations are formed.

We realize then that field correlations are intimately related to phase coherence. In fact, for nearly equal amplitudes, the normalized DNLSE field correlations ($\langle C_k \rangle / \langle C_0 \rangle$, see below the defining mathematical expressions) coincide with phase correlations ($\langle \cos \theta \rangle$), much like the phase coherence of a Bose–Einstein condensate in a lattice potential **[26]**.

The level of the formed field correlations can be measured. In trapped ultracold atoms experiments, the degree of phase coherence is directly related to the (measurable) visibility of fringes in the interference patterns formed after expansion of Bose–Einstein condensed atomic clouds **[27]**.

Positive Hamiltonian systems entertain two ground states (both with energy $\hbar_{a0}$), corresponding to the two values of signC (cf. Eq. **(3)**). Correlation distance ($k$) of the field functions of each of these two ground states extends to infinity. Absolute value of the normalized field correlations at all site-separations is unity ($|C_{kN}|_{ground\ state} = 1.0$) **[16]**.

Most interestingly, for high enough system nonlinearity, the field correlation function and the distribution of phases assume universal forms, independent of the exact value of the nonlinearity parameter and of system density. And equilibrium field correlations can be interrogated experimentally by the study of the equivalent optical system **[17]**.

The current "field correlations" section consists of two parts. In the first part we present field-correlations-related expressions. The second part is devoted to graphical illustrations.

*Expressions to calculate field correlations*. Correlation of fields separated $k$-sites apart, is defined as **[17]** -

$$C_k(z) = \frac{1}{N} \cdot \sum_{m=1}^{N} u_m(z) \cdot u_{m+k}(z) \cdot \cos[\theta_{m,k}(z)]$$

$$\theta_{m,k}(z) \equiv \phi_m(z) - \phi_{m+k}(z)$$

**(28)**

Let us first consider equilibrium nearest neighbor correlation.

It follows from Eqs. **(6)** and **(28)** that

$$C_0 = w_a \; ; \; C_1(z) = signC \cdot \frac{\mathcal{H}_2(z)}{2 \cdot N} = \frac{\hbar_2(z)}{2}$$

**(29)**

and

$$C_{1N}(z) \equiv \frac{C_1(z)}{C_0} = \frac{\hbar_2(z)}{2 \cdot w_a} = \frac{\hbar_a - \hbar_4(z)}{2 \cdot w_a}$$

**(30)**

Equation **(30)** holds for any distance ($z$), including of course long equilibrium distances. If the Lagrange parameters ($\beta, \mu$) are known, then the equilibrium site-averaged interaction energy $\left(\hbar_{4,eq}(\beta,\mu)\right)$ can be calculated as -

$$\hbar_{4,eq}(\beta,\mu) = \langle h_{4,eq}(\beta,\mu)\rangle = \frac{1}{2} \cdot \Gamma \cdot \int_0^\infty I^2 \cdot \mathcal{P}_I(\beta,\mu;I) \cdot dI$$

**(31)**

with $\mathcal{P}_I(\beta,\mu;I)$ given by Eq. **(11)**. Thus, at equilibrium, the value of nearest neighbors field correlation of DNLSE systems can be determined through finding the equilibrium interaction energy $\hbar_{4,eq}(\beta,\mu)$ (cf. equations **(30)**, **(31)**, and **(11)**).

Alternatively, if $C_{1N}$ is independently known, as prescribed for example in the next paragraph, then the values of both equilibrium tunneling energy $\left(\hbar_2(z_{equil.})\right)$ and equilibrium interaction energy $\left(\hbar_4(z_{equil.})\right)$ can be trivially determined.

A second, more general approach for calculating equilibrium field correlations at any site-separation is through the value of $\langle \cos\theta \rangle$, $\theta$ being fields' relative phase-angle. The Expression for field correlation at site-separation $k$ ($\langle C_k \rangle$) was derived by the authors of **[18]** as

$$\langle C_0 \rangle = \langle I \rangle = \langle w_a \rangle = w_a$$

$$\langle C_k \rangle = \langle \sqrt{I} \rangle^2 \cdot \langle \cos(\theta) \rangle^k \; ; \; k \geq 1$$

**(32)**

Elaborating on Eq. **(32)**, the authors of **[18]** say that the derived correlation equation is general, not limited to initial statistical-Gaussian excitation, and holds for field correlations in systems evolving under DNLSE dynamics for any long distance fields distribution where the *site-densities* ($I_m$'s) are not correlated and phase differences ($\theta_m$'s) are random (over realizations). Equation **(32)** shows that if the phase-differences are not flat-distributed (such that $\langle \cos(\theta) \rangle \neq 0$), then the fields are correlated and correlations exponentially decay with site-separation ($k$). If $\langle \cos(\theta) \rangle$ is negative, the sign of the fields' correlations alternates with $k$.

To determine $\langle \cos(\theta) \rangle$, the relative phase-angle density distribution $\left(\mathcal{P}_\theta(\theta)\right)$ must be known. An analytic expression for $\mathcal{P}_\theta(\theta)$ in the hot zone near the $L_i$ line - $\mathcal{P}_\theta(\theta) \propto \exp(2 \cdot \eta \cdot M_1^2 \cdot \cos\theta)$ – was derived in **[18]**, and an analytic expression for $\mathcal{P}_\theta(\theta)$ in the

- 23 -

entire cold zone - $\mathcal{P}_\theta(\theta) \propto \exp(2 \cdot \eta \cdot w_a \cdot \cos\theta)$ – was derived in **[19]**. The parameter $\eta$ is a Lagrange parameter with a value determined by solving an appropriate implicit equation. The parameter $M_1$ is a first-moment of an (assumed) initial statistical-Gaussian excitation. These PDFs*(θ)* were derived for strong system nonlinearity ($\Gamma \cdot w_a \gg 1$) justifying the quantum phase model approximation **[28]**,**[17]** under which the analytic PDFs$(\theta)$ expressions were derived.

Calculating $\langle \cos(\theta) \rangle$ according to the known PDFs$(\theta)$ and inserting in **(32)**, we arrive at an analytic expression for equilibrium field correlations, at any site-separation ($k$) for systems near and below the $L_i$ line on the strong nonlinearity portion of the DNLSE thermalization zone:

$$C_k = \langle C_k \rangle = \langle \sqrt{I} \rangle^2 \cdot \begin{Bmatrix} \left[(-1) \cdot \dfrac{\mathcal{I}_1(2 \cdot \eta \cdot w_a)}{\mathcal{I}_0(2 \cdot \eta \cdot w_a)}\right]^k & ; & \hbar_a(w_a) \le L_i(w_a) \\ \left[(-1) \cdot \dfrac{\mathcal{I}_1(2 \cdot \eta \cdot M_1^2)}{\mathcal{I}_0(2 \cdot \eta \cdot M_1^2)}\right]^k & ; & \hbar_a(w_a) > L_i(w_a) \end{Bmatrix} \quad ; \quad k \ge 1$$

$$C_{kN} \equiv \frac{\langle C_k \rangle}{\langle C_0 \rangle} = \frac{C_k}{w_a}$$

**(33)**

were $\mathcal{I}_n(z)$ is the modified Bessel function of the first kind.

The field correlation curves and map in the illustrating figures shown below were calculated either by direct running of system-evolution simulations and applying the definition **(28)**, or by executing the analytic expression **(33)**.

*Graphical illustrations*. The next four figures show field correlation values for systems on different parts of the DNLSE thermalization zone, along with graphical representations of the universality of field correlations formed during the evolution of DNLSE systems.

Start with ***Figure 14***. The red dots in all eight panels were calculated by evolution-to-equilibrium simulations and executing Eq. **(28)**. The red curves of all panels show high nearest-neighbors field correlations (absolute value of) for cold systems (between the vertical blue and vertical green lines) and essentially zero field correlations for hot systems close to the vertical red $\hbar_{a\infty}$ line. The blue dots were calculated analytically through Eq. **(33)**. The blue curves show that the analytic expression **(33)** predicts the value of the field correlations reasonably well for high absolute correlation values or at high system nonlinearity. Consulting the top three right panels and the bottom two right panels, we see that the red curves cross the green line at essentially $C_{1N} = -0.4$, indicating universality of DNLSE field correlations at high system nonlinearity **[17]**,**[18]**.

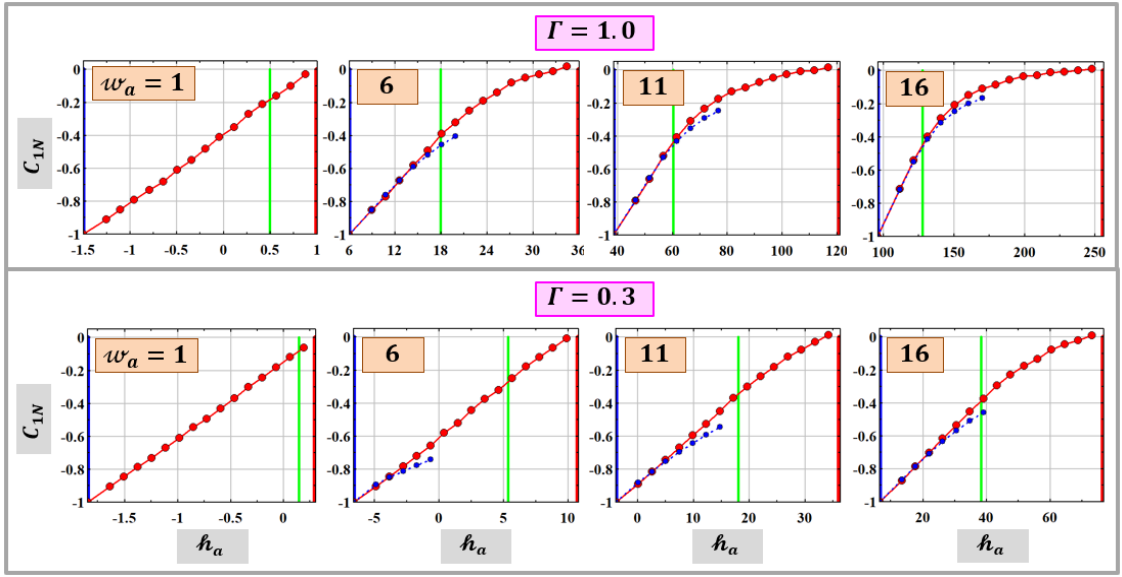

***Figure 14:*** *Normalized nearest-neighbors equilibrium field correlations ($C_{1N}$) as a function of site-averaged system energy at two different values of the nonlinearity parameter ($\Gamma$). The red dots were obtained by simulations (averaged over several realizations) through a focusing DNLSE equation (Eq.* **(3)**)*. The blue dots were calculated analytically (Eq.* **(33)**)*. The blue dots are missing from the two left panels as the quantum phase approximation is invalid for such low system nonlinearities ($\Gamma \cdot w_a \sim 1$). Colored vertical lines as in* ***Figure 4****. Clearly, normalized field correlations at all parameters are maximal (in absolute value) ($C_{1N} = -1$) at the ground state, and are all monotonically reduced as system energy (and system temperature) goes up, reaching a zero value at the infinite temperature line.*

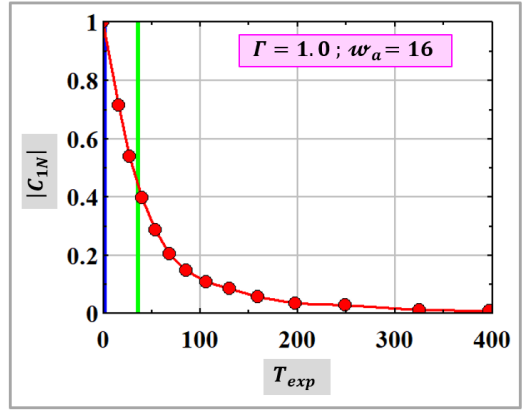

***Figure 15:*** *Absolute value of nearest-neighbors field correlation vs. system temperature. The red dots were calculated through evolution-to-equilibrium simulations and executing Eq.* **(28)**. *System density is taken to be constant at $w_a = 16.0$. System energy ($\hbar_a$) is a parameter, i.e. the data points of the curve are on the plane $\left(|C_{1N}(\hbar_a)|, T_{exp}(\hbar_a)\right)$ with system energy going straight up (on the DNLSE phase diagram) from the $\hbar_{a0}$ line towards the $\hbar_{0\infty}$ line. The vertical green line designates the temperature of a system on the $L_i$ line. The figure clearly shows the inverse relations between field correlation and system temperature. Explicitly – higher system temperature lower absolute value of the system's field correlation.*

The curve of *Figure 15* indicates the inverse relations of system temperature and field correlation (in absolute value). Explicitly – higher system temperature, weaker field correlations. Remembering that temperature is proportional to the width of PDF($I$), we can make a parallel inverse-relation statement: the width of equilibrium PDF($I$) is inversely related to field correlation. Explicitly: wider PDF($I$), weaker field correlation.

Next, *Figure 16* shows the remarkable universality property of equilibrium field correlations formed during evolution of DNLSE systems at high system nonlinearities ($\Gamma \cdot w_a \gg 1$). The straight horizontal lines of *Figure 16*, each of a different energy, extend the correlation's universality findings of **[17]**, and of **[18]**. The data points of the figure were calculated through Eq. **(33)**. As shown very clearly, at high system nonlinearity, normalized equilibrium values of field correlations of all DNLSE systems with energies $\hbar_a(q_g; w_a) = \hbar_{a0}(w_a) + q_g \cdot 2 \cdot w_a$ are practically equal, independent of the value of the nonlinearity parameter ($\Gamma$) and independent of system density ($w_a$). This correlation-level equality extends to each and every site-separation ($k$).

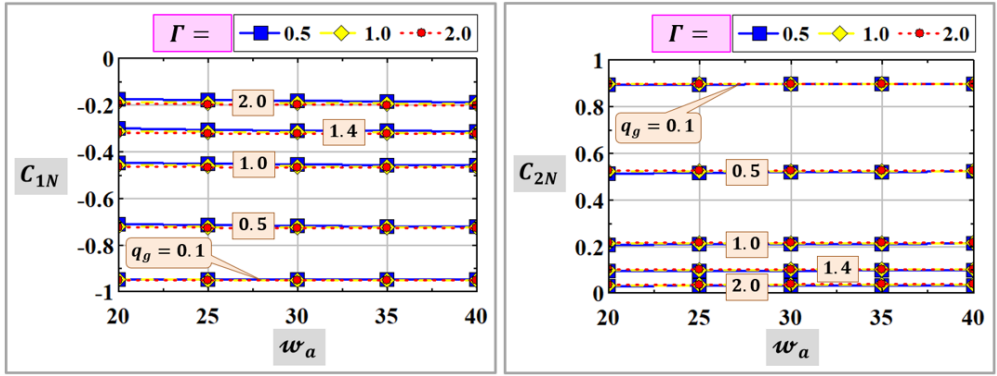

*Figure 16:* *Universality of the DNLSE equilibrium field correlations. Shown are normalized values (Eq. **(33)**) of the nearest-neighbors ($C_{1N}(z_{equil})$ – left) and next-nearest-neighbors ($C_{2N}(z_{equil})$ - right) field correlations as a function of the site-averaged density ($w_a$) and the nonlinearity parameter ($\Gamma = 0.5, 1.0, 2.0$). The parameter ($q_g$) is a multiplier of the energy distance between the ground-state and the $L_i$ line (i.e. $\hbar_a(q_g; w_a) = \hbar_{a0}(w_a) + q_g \cdot 2 \cdot w_a$). The remarkable universality characteristics of the field correlations in equilibrated DNLSE systems is clearly revealed. Namely, at strong enough system nonlinearity ($\Gamma \cdot w_a \gg 1$), the values of every site-separation ($k$) field correlation is independent of system density and of the nonlinearity parameter ($\Gamma$) at a fixed $q_g$ energy level across the entire DNLSE thermalization zone. Along the $L_i$ line for example ($q_g = 1.0$), the value of $C_{1N}$ is $\pm \approx$ 0.455, and the value of $C_{2N}$ is $+0.215$, independent of the value of the nonlinearity parameter ($\Gamma$). The shown several straight horizontal correlation lines of each panel extend the correlation's universality findings of **[17]**, and of **[18]**.*

The next figure - *Figure 17* – compares the characteristics of a linear system ($\Gamma = 0$) and the characteristics of a nonlinear DNLSE system. Both systems are initially excited onto the lower part of the cold zone, near the $\hbar_{a0}$ line (panels D,H). Key differences are seen in the PDFs($I$), panels A,E and in the constant vs. decaying field correlations in panels C,G.

In the extreme case of system nonlinearity approaching zero (essentially by $\Gamma \to 0$), system temperatures rise to infinity everywhere on the triangular-shaped phase diagram, with the exception of the zero-variance non-energy-diffusing ground state. Stated differently: a DNLSE system above ground state is formally approaching an infinite temperature as the nonlinearity parameter is approaching zero, yet except for systems on the $L_i$ line (cf. *Figure 7*), the PDFs($I$) are of a finite variance (panel A of *Figure 17*).

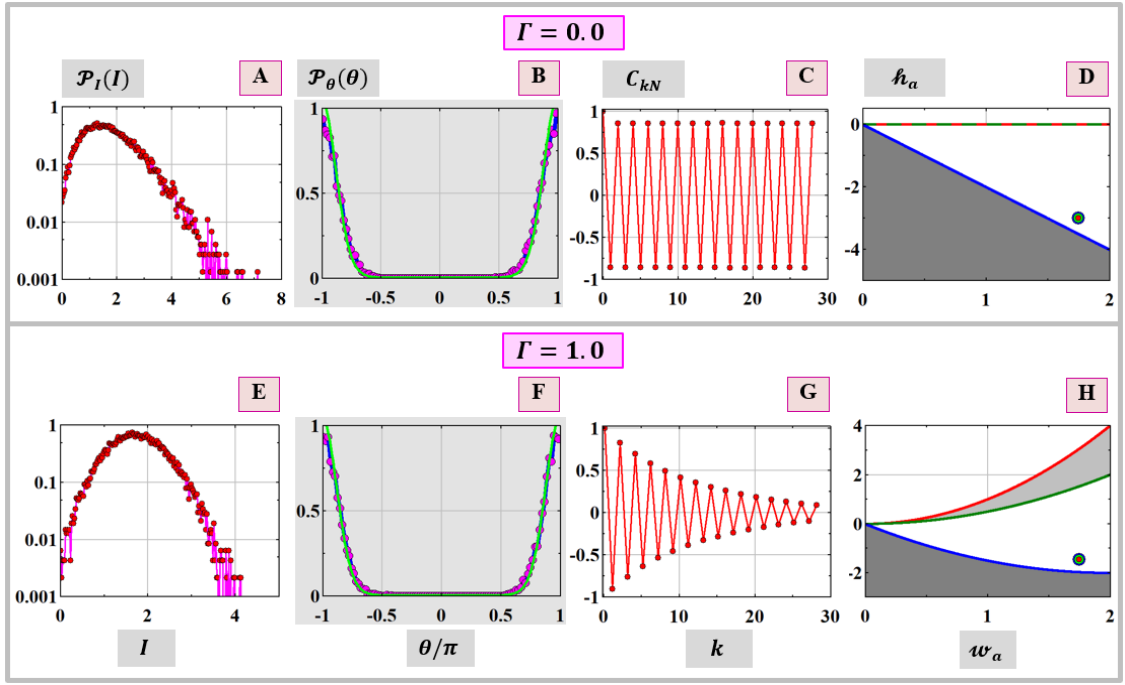

*Figure 17: Characteristics of an equilibrated linear system (top row, $\Gamma = 0$) vs. an equilibrated nonlinear system (bottom row, $\Gamma = 1.0$). A,E: Simulated PDF($I$). As shown, the shape of the $\mathcal{P}_I(I)$ curve for the linear system does not resemble a Gaussian at all. Note the high amplitude oscillators of the stationary-statistics-reached linear system (A). B,F: PDF($\theta$). The magenta dots are simulated. The continuous green curve is theoretical [19]. In both linear-nonlinear cases, relative phase-angles $(\theta_m = \phi_m - \phi_{m+1})$ are narrowly concentrated around $\pi$, as expected for systems initially excited close to the ground state and evolving under the dynamics a focusing-type DNSLE equation. C,G: Simulated field correlations. Considering the linear system, if the system is initially excited with uniform amplitudes and with a small random-phase window, the launched fields are strongly correlated, independent of site-separation ($k$). These k-independent strong correlations are maintained during evolution of the single-term-Hamiltonian linear system (C). Unlike field correlations of the linear system, during evolution of the two-term-Hamiltonian nonlinear system ($\Gamma > 0$), the initially-formed strong field correlations do gradually weaken (for $k \geq 2$) to reach a k-dependent exponential decay at steady state distances (independent of nonlinearity value) (panel G). D,H: System position on the DNLSE phase diagram ($w_a = 1.75$). Looking at the panel for the linear system (D), note the coincidence of the infinite temperature line ($\hbar_{a\infty}$) with the intermediate ($L_i$) line, forming a triangular-shaped thermalization zone. In terms of temperature, the equilibrated linear*

*system (and all other systems in the linear thermalization zone, except for ground-state-excited systems) is formally at infinite temperature. In contrast, the nonlinear system, initially excited near the ground state line is quite "cold".*

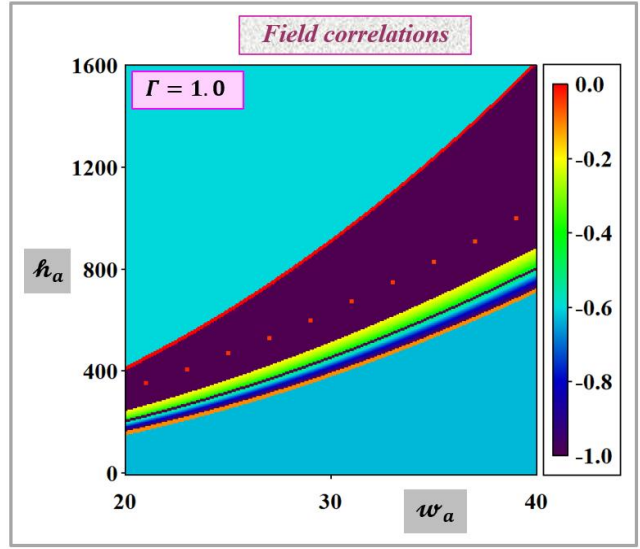

*Figure 18: Map of DNLSE equilibrium nearest neighbors field correlations ($C_{1N}$) on a strong system nonlinearity portion ($\Gamma \cdot w_a \gg 1$) of the thermalization zone. Same-color points are all at a distance of $q_g$ (i.e. at a distance of $q_g \cdot 2 \cdot w_a$) from the $\hbar_{a0}$ line (or "the ground-state line", designated by the bottom orange line). The black line crossing the colored area is at a distance of $q_g = 1.0$. The yellow line at the top of the colored portion is at a distance of $q_g = 2.0$, showing a value of $C_{1N} = -0.2$ (cf.* ***Figure 16****). Thus, here too, universality of the DNLSE equilibrium field correlations is clearly revealed. The colored portion was calculated by the analytic expression (Eq.* **(33)***). The black area was left untouched. The ten isolated discrete nearly-red-colored dots are at a distance of $q_g = 4.0$. Their $C_{1N}$ values, ranging between $-0.058$ and $-0.027$, were determined by field-propagation simulations (averaged over several realizations). The map area above these discrete point should have been covered entirely by a red color. Namely - essentially zero field correlations at this high temperature region of the DNLSE phase diagram.*

The last figure of this section - ***Figure 18***, is a map of normalized equilibrium field correlations. The map shows a relatively strong nonlinearity region of the thermalization zone. Values for the colored part were determined analytically (Eq. **(33)**). Values for the discrete red dots by evolution-to-equilibrium simulations. The map area above these discrete point should have been covered red, signifying no correlations. Looking at the map, two effects become clear. First, once more – the universality of DNLSE field correlations at high system nonlinearity as discussed above. Second - the fact of essentially no field correlations at the upper region of the hot zone. Namely - no field correlations at a high temperature-high system nonlinearity region of the DNLSE phase diagram.

In the next section, along with a complementary appendix (***Appendix 1***), we derive an approximate analytic expression to the grand-canonical partition function (Eq. **(9)**) and to the associated distribution function of site densities (PDF($I$) – Eq. **(11)**). We show

## 8. Analytic PDF(I) expression for systems on the hot zone

The symmetric nonnegative kernel related to the grand canonical partition function of large DNLSE systems is the product of two decaying exponential functions and a zero order modified Bessel function of the first kind (Eq. **(9)**). In the "second-order" approximation, we represent the modified Bessel function by the first two terms of its series expansion (cf. *Appendix 1* below). Eigenvalues ($\lambda_i$) and nonnegative eigenfunctions ($v_i(x)$) of this kernel are the solutions of the Fredholm integral equation of the second kind with symmetric kernel **[29]**:

$$\int_0^\infty [F(x) \cdot G(y) \cdot (1 + \beta^2 \cdot x \cdot y)] \cdot v_i(x) \cdot dx = \lambda_i \cdot v_i(y)$$

$$F(x) \equiv e^{\left[-\beta \cdot \left(\frac{\mu}{2} \cdot x + \frac{\Gamma}{4} \cdot x^2\right)\right]} \quad ; \quad G(y) \equiv e^{\left[-\beta \cdot \left(\frac{\mu}{2} \cdot y + \frac{\Gamma}{4} \cdot y^2\right)\right]}$$

(34)

with the largest eigenvalue ($\lambda_1$) directly related to the partition function (Eq. **(10)**) and the associated eigenfunction ($v_1(x)$) directly related to the PDF($I$) of the equilibrated DNLSE system in question (Eq. **(11)**).

It turns out that if the Bessel function of equation **(9)** is described by a "second-order" approximation (for arguments smaller than unity), namely by taking the first two terms of the otherwise infinite series describing the function, then the Fredholm integral equation **(34)** can be solved analytically (cf. *Appendix 1* below).

The solved expressions for the integral equation with a second-order approximated Bessel function (Eq. **(34)**) are (*Appendix 1*):

$$\lambda_{1E}(\beta, \mu) = \frac{e^{q_0^2} \cdot v_0}{2 \cdot \sqrt{\epsilon}} + \kappa_0 \cdot \frac{1 - e^{q_0^2} \cdot q_0 \cdot v_0}{2 \cdot \epsilon}$$

$$\mathcal{P}_{IE}(\beta, \mu; I) = C^{-1} \cdot e^{-\left(\sqrt{\epsilon} \cdot I + q_0\right)^2} \cdot (1 + \kappa_0 \cdot I)^2$$

(35)

where $\lambda_{1E}$ is the largest eigenvalue and $\mathcal{P}_{IE}(\beta, \mu; I) = v_{IE}^2(\beta, \mu; I)$ where $v_{IE}(\beta, \mu; I)$ is the associated eigenfunction. All constants of Eq. **(35)** are defined in *Appendix 1*.

Consulting Eq. **(35)**, we see that the second-order PDF($I$) is a product of a Gaussian and a parabola. The *parabolic-parameter* $\kappa_0$ appearing in the parabolic-term is smaller than the value of the Lagrange parameter $\beta$ and is thus smaller than unity if $\beta$ is smaller than unity, i.e. if system temperature is greater than the value of $1/\Gamma$.

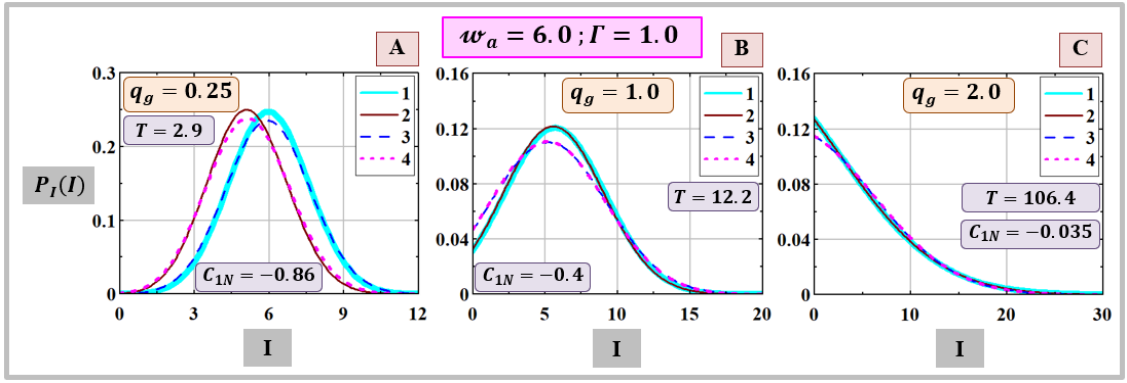

*Figure 19:* *Distribution functions of site densities* $(\mathcal{P}_I(I))$. *Four curves are shown in each panel – the continuous cyan curve (1) and the dashed blue curve (3) show the* $\mathcal{P}_I(I)$ *of Eq.* **(11)**, *calculated with all Bessel terms in place. The continuous thin-dark curve (2) and the dashed magenta curve (4) show the approximate analytic* $\mathcal{P}_{IE}(I)$ *of Eq.* **(35)**, *calculated with only two Bessel terms in place. The continuous cyan curve (1) and the continuous thin-dark curve (2) were calculated with exact* $(\beta, \mu)$ *values determined by numeric inversion of Eq.* **(15)**. *The* $(\beta, \mu)$ *values for the two dashed curves (3 and 4) are actually* $(\beta_{exp}, \mu_{exp})$ *determined through the analytic expression* **(25)** *using* **(24)** *and through the implicit integral equation* **(26)**. *The panels show that second-order PDF(I) of the "cold" system (A) does not match the exact PDF(I). However, the other two panels (B,C) show that second-order PDFs(I) of Eq.* **(35)** *do match the exact PDFs(I) for the two higher energy systems.*

The three panels of *Figure 19* show the equilibrium PDF($I$) of DNLSE systems at three different energies: $\hbar_a = $ "cold" ; $\hbar_a = L_i$ ; $\hbar_a = $ "hot", all at the same density of $w_a = 6.0$. System nonlinearity is also fixed at $\Gamma \cdot w_a = 6.0$. The parabolic-parameter $\kappa_0$ values for the (A;B;C) panels of *Figure 19* are $\kappa_0 = $ (0.575 ; 0.038 ; 0.00057) respectively. The panels show that second-order PDF($I$) of the "cold" system does not match the exact PDF($I$). However, the other two panels show that second-order PDFs($I$) of Eq. **(35)** do match the exact PDFs($I$) for the two higher energy systems.

Consulting Eq. **(35)** and the panels of *Figure 19*, we can make the following observations:

- Given high system nonlinearity ($\Gamma \cdot w_a \gg 1$), The equilibrium PDFs($I$) of systems at or above the $L_i$ line are well described by the analytic PDF($I$) expression of Eq. **(35)**.

- As the parabolic curves in the analytic PDF($I$) expression (Eq. **(35)**) for the high energy systems are very shallow, the Gaussian-parabola product is another slightly shifted Gaussian with essentially the same width. Thus, at high system nonlinearity, the shape of the equilibrium PDF($I$) curves for DNLSE systems at or above the $L_i$ line is practically a pure (possibly truncated) Gaussian shape.

- The parameter $\epsilon$ in the $\mathcal{P}_{IE}(\beta, \mu; I)$ expression of Eq. **(35)** is related to the (untruncated) Gaussian variance $\sigma^2$ as $\epsilon^{-1} = 2 \cdot \sigma^2$. Thus, looking at the definition of $\epsilon$ in Eq. **(39)** of *Appendix 1* $\left(\epsilon \equiv \frac{1}{2} \cdot \beta \cdot \Gamma\right)$ we find: $T = (\beta \cdot \Gamma)^{-1} = \sigma^2$. Put it all together: given high system nonlinearity, the shape of the equilibrium

PDF(I) curves for DNLSE systems at or above the $L_i$ line is practically a pure (possibly truncated) Gaussian shape with `variance` ($\sigma^2$) equal to the temperature of the system (Eq. **(24)**).

These "second-order-related" observations end the present analytic PDF($I$) section and actually end our entire detailed characterization of equilibrated large DNLSE systems.

## 9. Summary

The subject of the study presented here is a system consists of a long $1d$ array of coupled nonlinear oscillators. The evolution dynamics of the system is taken to be the DNLSE (Eq. **(3)**). A DNLSE system initially excited onto a point ($w_a, \hbar_a$) of a well-defined thermalization zone of the DNLSE phase diagram (Eq. **(2)**) will thermalize **[7]** (cf. *Figure 3*). The equilibrium statistical properties of a thermalized system are predictable.

Previous studies predicted equilibrium PDFs and temperatures for strong system nonlinearities by system-entropy separation into the sum of density-entropy and a relative-phase entropy. Entropy separation was justified in the strong system nonlinearity regime by the quantum phase approximation. Entropy separation led to the derivation of system temperatures as well as the derivation of analytic expressions for equilibrium PDF($I$) and equilibrium PDF($\theta$) **[17]**-**[19]**.

Here we have taken a rigorous approach of predicting system equilibrium statistics based on the thermodynamics formalism of grand canonical ensembles (cf. *Figure 2* and related text). The thermodynamics approach allowed the extension of system-characteristics predictions to cover the entire thermalization zone.

The grand canonical statistics is determined by two Lagrange parameters - $(\beta, \mu)$. To determine the values of $\beta$ and $\mu$ we have taken here one of two routes. The first is numerically inverting the two exact thermodynamics partial derivatives of Eq. **(15)**. Such 2D numerical inversion is challenging and could sometimes introduce uncertainty errors. The second is finding $\beta_{exp}$ through execution of the analytic expression **(25)**, along with expression **(24)**. We have derived the analytic expression **(25)** from an approximate analytic expression of system entropy published in **[9]**. Once the value of $\beta_{exp}$ is determined, the value of $\mu_{exp}$ is calculated through solving the *one parameter* implicit integral equation **(26)**. Once $\left(\beta(w_a, \hbar_a), \mu(w_a, \hbar_a)\right)$ are determined, equilibrium statistical properties of the system studied such as entropy, temperature, chemical potential, nearest-neighbors field correlations, and PDF($I$) can be calculated.

To derive physics insights from numerical calculations, graphical representations are typically required. Analytic expressions on the other hand often provide deeper physics understanding by direct inspection. As part of the present study we have gone through deriving an approximate analytic PDF($I$) expression (section **8** and *Appendix 1*). Probing further the derived expression enabled the insightful identification of system's temperature with the variance of the system's PDF($I$) (cf. the last part of section **8**).

Crossing the DNLSE thermalization zone is an intermediate $L_i$ line (Eq. **(2)**). The $L_i$ line divides the thermalization zone into a cold zone below the line and a hot zone above the line (**[19]**). Characteristics of DNLSE systems can be generalized in relations to these cold-hot zones.

Some of our deduced characteristics of equilibrated DNLSE systems are the following:

- Temperatures of systems on the cold zone rise approximately linearly with energy. Temperatures of systems on the hot zone rise exponentially-like with energy.
- On the high system nonlinearity region of the thermalization zone ($\Gamma \cdot w_a \gg 1$) the PDF($I$) is of a Gaussian shape, possibly truncated.
- On the high system nonlinearity region of the thermalization zone, system temperature and the `variance` of its equilibrium PDF($I$) are equal.
- Chemical potentials of systems on the hot zone rise exponentially-like with energy.
- Field correlations are inversely related to system temperature - higher system temperature, lower absolute value of the system's field correlation.
- Field correlations for all site-separations ($C_{kN}$) of systems on the high system nonlinearity region of the thermalization zone at all energies are universal – their values are independent of system density or of the value of the nonlinearity parameter. At energies of $4 \cdot w_a$ above ground state, absolute value of nearest-neighbors normalized field correlation is already as low as $0.2$ (cf. *Figure 18*).
- On the high system nonlinearity region of the thermalization zone, nearest-neighbors field correlations of DNLSE systems with energies ($\hbar_a$) of $8 \cdot w_a$ or more above ground state practically vanish (cf. *Figure 18*).
- Given high system nonlinearity, the equilibrium PDF($I$) of systems at or above the $L_i$ line are well described by the analytic PDF($I$) expression of Eq. **(35)**.

These observations, derived for an abstract DNLSE system, hold of course for the equilibrium properties of the variety of actual discrete nonlinear physical systems with evolution dynamics approximated by a DNLSE type equation as analyzed in the present study.

**Acknowledgement**

I would like to thank Prof. David Mukamel for his many guiding comments and constructive suggestions during preparation of the manuscript.

*Appendix 1:* **Approximate distribution function of site densities**

In this appendix we derive an approximate analytic expression for the grand canonical partition function of Eq. **(9)** along with an approximate analytic expression for the distribution function of site densities $(\mathcal{P}_I(I))$. We derive the approximate expressions through taking only two first terms of the infinite series by which the zero order modified Bessel function of the first kind is described. We refer to this approximation as a "second-order" approximation.

The infinite series describing the zero order modified Bessel function of the first kind ($\mathcal{I}_0(z)$, Eq. **(13)**) with $z$ a complex number, is given by the expression **[30]**:

$$\mathcal{I}_0(z) = \sum_{k=0}^{\infty} \frac{\left(\frac{1}{4} \cdot z^2\right)^k}{(k!)^2}$$

(36)

If the argument of the modified Bessel function is small $\left(2 \cdot \beta \cdot \sqrt{x \cdot y} < 1\right)$ for $(x, y)$ in the range of interest (determined by the decay-rate of the preceding exponential functions), then the modified Bessel function can be approximated by the first two terms:

$$\mathcal{I}_0\left(2 \cdot \beta \cdot \sqrt{x \cdot y}\right) \cong 1 + \beta^2 \cdot x \cdot y \ ; \ 2 \cdot \beta \cdot \sqrt{x \cdot y} < 1$$

(37)

The kernel of Eq. **(13)** is approximated then as $\mathcal{K}(x, y) \cong F(x) \cdot G(y) \cdot (1 + \beta^2 \cdot x \cdot y)$ with $F(x) \equiv e^{\left[-\beta \cdot \left(\frac{\mu}{2} \cdot x + \frac{\Gamma}{4} \cdot x^2\right)\right]}$; $G(y) \equiv e^{\left[-\beta \cdot \left(\frac{\mu}{2} \cdot y + \frac{\Gamma}{4} \cdot y^2\right)\right]}$. We are now looking for an eigenvalue $(\lambda_1)$ and an eigenfunction $(v_1(x))$ such that **[29]**:

$$\int_0^{\infty} [F(x) \cdot G(y) \cdot (1 + \beta^2 \cdot x \cdot y)] \cdot v_{1E}(x) \cdot dx = \lambda_{1E} \cdot v_{1E}(y)$$

(38)

The subscript "$E$" in **(38)** stands for "expression".

It is convenient at this point to define a set of four parameters $(\epsilon; \alpha; q_0; v_0)$ each depends, directly or indirectly, on the two Lagrange parameters $(\beta, \mu)$ as:

$$\epsilon \equiv \frac{1}{2} \cdot \beta \cdot \Gamma \ ; \ \alpha \equiv \beta \cdot \mu \ ; \ q_0 \equiv \frac{\alpha}{2 \cdot \sqrt{\epsilon}} \ ; \ v_0 \equiv \sqrt{\pi} \cdot erfc(q_0)$$

(39)

Working out the lengthy mathematics, we first determine the value of a *parabolic-parameter* - $\kappa_0$ as:

$$\kappa_0^\pm = \frac{-b \pm \sqrt{b^2 + 4 \cdot \beta^2}}{2} \; ; \; \kappa_0 \equiv \kappa_0^+ \; ; \; b \equiv \frac{I_0 - \beta^2 \cdot I_2}{I_1}$$

$$I_0 = \frac{v_0}{2 \cdot \sqrt{\epsilon}} \; ; \; I_1 = \frac{e^{-q_0^2} - q_0 \cdot v_0}{2 \cdot \epsilon}$$

$$I_2 = \frac{-2 \cdot e^{-q_0^2} \cdot q_0 + (1 + 2 \cdot q_0^2) \cdot v_0}{4 \cdot \epsilon^{3/2}}$$

**(40)**

And then, with $\kappa_0$ in place, calculate the thought-for expressions for $\big(\mathcal{Z}(\beta,\mu); \lambda_1(\beta,\mu); \mathcal{P}_I(\beta,\mu;I)\big)$ as:

$$\mathcal{Z}_E(\beta,\mu) \cong [2 \cdot \pi \cdot \lambda_{1E}(\beta,\mu)]^N$$

$$\lambda_{1E}(\beta,\mu) = \frac{e^{q_0^2} \cdot v_0}{2 \cdot \sqrt{\epsilon}} + \kappa_0 \cdot \frac{1 - e^{q_0^2} \cdot q_0 \cdot v_0}{2 \cdot \epsilon}$$

$$\mathcal{P}_{IE}(\beta,\mu;I) = \mathcal{C}^{-1} \cdot e^{-(\sqrt{\epsilon} \cdot I + q_0)^2} \cdot (1 + \kappa_0 \cdot I)^2$$

$$\mathcal{C} = I_0 + 2 \cdot \kappa_0 \cdot I_1 + \kappa_0^2 \cdot I_2$$

**(41)**

where $\lambda_{1E}(\beta,\mu)$ and $v_{1E}(\beta,\mu;I)$ in $\mathcal{P}_{IE}(\beta,\mu;I) = v_{1E}^2(\beta,\mu;I)$ are the solutions of Eq. **(38)**.

As $\mathcal{P}_{IE}(I)$ is explicitly known, the statistical averages of system density $\big(w_{aE}(\beta,\mu)\big)$, equilibrium interaction energy $\big(\hbar_{4E}(\beta,\mu)\big)$, and equilibrium tunneling energy $\big(\hbar_{2E}(\beta,\mu)\big)$ can be explicitly known as well:

$$I3 = \frac{e^{-q_0^2} \cdot (1 + q_0^2) - q_0 \cdot (1.5 + q_0^2) \cdot v_0}{2 \cdot \epsilon^2}$$

$$I4 = \frac{-e^{-q_0^2} \cdot 2 \cdot q_0 \cdot (5 + 2 \cdot q_0^2) + (3 + 12 \cdot q_0^2 + 4 \cdot q_0^4) \cdot v_0}{8 \cdot \epsilon^{5/2}}$$

$$w_{aE}(\beta,\mu) = \mathcal{C}^{-1} \cdot (I_1 + 2 \cdot \kappa_0 \cdot I_2 + \kappa_0^2 \cdot I_3)$$

$$\hbar_{4E}(\beta,\mu) = \frac{\Gamma}{2} \cdot \mathcal{C}^{-1} \cdot (I_2 + 2 \cdot \kappa_0 \cdot I_3 + \kappa_0^2 \cdot I_4)$$

$$\hbar_{2E}(\beta,\mu) = \hbar_a - \hbar_{4E}(\beta,\mu)$$

**(42)**

As shown by *Figure 19* in the main text, the derived $\mathcal{P}_{IE}(I)$ expression of Eq. **(41)** well approximates the exact $\text{PDF}(I)$ of Eq. **(11)** on a large region of the DNLSE thermalization zone.